\documentclass[journal]{IEEEtran}
\usepackage{enumitem}
\usepackage{float}
\usepackage{graphicx}
\usepackage{enumerate}
\usepackage[fleqn]{amsmath}
\usepackage{color}
\usepackage{threeparttable}
\usepackage{stfloats}
\usepackage[linesnumbered,ruled,vlined]{algorithm2e}
\usepackage{setspace}

\usepackage{subfigure}
\usepackage{graphicx}
\usepackage{subfig}
\usepackage{caption}
\usepackage{enumitem}
\setlist[itemize]{leftmargin=2em}
\usepackage{enumitem}
\setlist[itemize]{leftmargin=2em}
\graphicspath{{figures/}{pictures/}{images/}{./}}
\SetKwRepeat{FunctionPruning}{Function \textsc{Pruning}($W$)}{}
\SetKwRepeat{FunctionQuantizing}{Function \textsc{Quantizing}($W$)}{}
\SetKwRepeat{FunctionProcedure}{\textsc{Procedure}()}{}

\setenumerate[1]{itemsep=0pt,partopsep=0pt,parsep=\parskip,topsep=5pt}
\setitemize[1]{itemsep=0pt,partopsep=0pt,parsep=\parskip,topsep=5pt}
\setdescription{itemsep=0pt,partopsep=0pt,parsep=\parskip,topsep=5pt}

\ifCLASSINFOpdf
\else
\fi

\begin{document}
\label{Title}
\title{ISCom: Interest-aware Semantic Communication Scheme for Point Cloud Video Streaming
\vspace{0.2em}}

\label{Author}
\author{
	Yakun~Huang$^{\dagger}$,
	Boyuan~Bai$^{\dagger}$,
	Yuanwei~Zhu,
	Xiuquan~Qiao$^{\star}$,
	Xiang~Su~\IEEEmembership{member,~IEEE},
	and~Ping~Zhang~\IEEEmembership{Fellow,~IEEE}
	\IEEEcompsocitemizethanks{
		\IEEEcompsocthanksitem  Y. Huang, B. Bai, Y. Zhu, X. Qiao, and P. Zhang are with State Key Laboratory of Networking and Switching Technology, Beijing University of Posts and Telecommunications, Beijing 100876, China. E-mail:\{ykhuang, baiboyuan, zhuyw, qiaoxq, pzhang\}@bupt.edu.cn.
	    \IEEEcompsocthanksitem X. Su is with the Department of Computer Science, Norwegian University of Science and Technology, 2815 Gj{\o}vik, Norway and University of Oulu, 90570, Oulu, Finland. Email:xiang.su@ntnu.no.
	    \IEEEcompsocthanksitem $^{\dagger}$Both authors contributed equally to this work. $^{\star}$Xiuquan Qiao is the corresponding author.
}
\vspace{-1.7em}}

\markboth{Journal of \LaTeX\ Class Files,~Vol.~XX, No.~X, XX~2020}%
{Shell \MakeLowercase{\textit{et al.}}: Bare Demo of IEEEtran.cls for IEEE Journals}

\maketitle
\label{Abstract}
\begin{abstract}
The provisioning of immersive point cloud video (PCV) streaming on pervasive mobile devices is a cornerstone for enabling immersive communication and interactions in the future 6G metaverse era. However, most streaming techniques are dedicated to efficient PCV compression and codec extending from traditional 3-DoF video services. Some emerging AI-enabled approaches are still in their infancy phase and are constrained by intensive computational and adaptive flow techniques. In this paper, we present ISCom, an Interest-aware Semantic Communication Scheme for PCV, consisting of a region-of-interest (ROI) selection module, a lightweight PCV streaming module, and an intelligent scheduler. 
First, we propose a two-stage efficient ROI selection method for providing interest-aware PCV streaming, which significantly reduces the data volume. Second, we design a lightweight PCV encoder-decoder network for resource-constrained devices, adapting to the heterogeneous computing capabilities of terminals. Third, we train a deep reinforcement learning (DRL)-based scheduler to adapt an optimal encoder-decoder network for various devices, considering the dynamic network environments and computing capabilities of different devices. Extensive experiments show that ISCom outperforms baselines on mobile devices at least 10~FPS and up to 22~FPS.
\end{abstract}


\IEEEpeerreviewmaketitle

\section{Introduction}
\IEEEPARstart{P}{oint} cloud video (PCV), represented as 3D unordered points and RGB color or mesh format, adds an additional 3-DoF movement to VR and panoramic videos than traditional 3-DoF rotation.
PCV enables immersive and interactive experiences and has become a cornerstone for digital twins and future metaverse~\cite{cai2022towards}.
Streaming massive point cloud video in real time puts new demands on existing network infrastructure and video codecs, including: 1) Ultra-high bandwidth and ultra-low delay. PCV services require at least 1 Gbps bandwidth and reach 1 Tbps level with thousands of concurrent streams. In addition, the ideal delay is less than 5ms and is more stringent than that of traditional 3-DoF videos (i.e.,$<$20 ms)~\cite{strinati20196g}.
2) Heavy computational requirements for codec. Encoding and decoding a point cloud video using the MPEG standard are already computation intensive.
For instance, encoding a one-second long video from the longdress dataset with lossy compression requires 11 to 42 minutes using MPEG V-PCC on a generic computer~\cite{lee2020groot}. Thus, massive codec computation hinders the provision of a 6-DoF experience on mobile devices, such as Augmented Reality (AR) and Virtual Reality (VR) glasses.

Existing research providing point cloud video delivery can be classified into two categories, i.e., extending 3-DoF video streams and AI-driven techniques. Approaches in the first category explore the PCV compression and codec methods to boost the user experience, such as Google Draco~\cite{draco}, Oct-tree~\cite{schnabel2006octree}, MPEG V-PCC/G-PCC~\cite{graziosi2020overview}, and some deep learning-based compression and encoding~\cite{huang20193d,zhang2020mobile}.
Besides, the extended 3-DoF tiling-based approaches~\cite{li2020joint,liu2021point,subramanyam2020user,li2022optimal} of predicting the user perspectives that combines together compression, this can then reduce the transmission data volume, which leads to an improvement in the efficiency and user experience.
Recently, the second approach, i.e., AI-native semantic communication, has introduced new ideas to address the massive PCV transmission challenge.
This approach reduces the transmission data volume by extracting key features of video frames and reconstructing video frames on the device side, reducing the demand for network bandwidth.
Xie \textit{et al.}~\cite{xie2021deep} proposed a deep learning-based semantic communication system for text transmission. They extended it into a multi-user scenario to execute the visual question-answering task.
Bourtsoulatze \textit{et al.}~\cite{bourtsoulatze2019deep} proposed joint source-channel coding motivated semantic communication for image transmission tasks.
AITransfer~\cite{huang2021aitransfer} provided a semantic-aware transmission for real-time volumetric video services and explored an adaptive streaming technique for point cloud video.
Although these deep learning-based methods can efficiently extract semantic features, they are computation intensive and difficult to deploy on resource-constrained devices, reducing the network resources by increasing the computation. Thus, AI-driven approaches are still in the exploratory stage.

Aiming to achieve a real-time PCV service on resource-constrained devices, there exist a number of key challenges, including,
\begin{itemize}[leftmargin=0.6cm,topsep=0.1cm]
	\setlength{\itemsep}{2pt}
	\setlength{\parsep}{0pt}
	\setlength{\parskip}{0pt}
	\item \textbf{Complex, redundant PCV content greatly increases the transmission volume and computational overhead.}
	PCV shares similar viewing characteristics to 3-DoF video in that the user's view range is limited at every single moment. 
	Hence, dynamically selecting the ROI for targeted transmission is an important technique to enhance PCV transmission efficiency and reduce energy consumption.
	However, existing 3-DoF video user viewpoint prediction still lacks the interest perception to integrate dynamic and static multidimensional features for selection. 
	Thus, selecting ROI from the complex PCV frames with existing 3-DoF methods is challenging.
	
	\item \textbf{Computation-intensive AI decoding hinders real-time PCV streaming on pervasive devices.}
	Emerging AI-driven transmission and semantic communication technologies promises to achieve real-time PCV transmission (i.e., at least 10~FPS).
	Although existing AI-driven transmission and applications concentrate on text, audio, and 2D video, implementing real-time decoding on mobile devices for 6-DoF PCV is unrealistic (e.g., only 1-2~FPS on ubiquitous smartphones). 
	The massive computation required by AI inference challenges us to implement real-time PCV delivery on resource-constrained devices.
	
	\item \textbf{Heterogeneous interaction devices with various computing capabilities challenge the adaptive transmission of AI-driven PCV.}
	AI-driven PCV streaming is a promising solution for implementing low bandwidth consumption under existing network infrastructure. 
	To enable the adaptive streaming for AI-driven PCV, dynamic network environments and heterogeneous pervasive devices with uneven computing capabilities affect the AI-native and semantic communication-based PCV transmission.
	This also illustrates that providing real-time PCV transmission service for these pervasive devices adaptively and intelligently is a significant and challenging effort.
\end{itemize}

We contribute, ISCom, an innovative interest-aware semantic communication scheme for PCV streaming, consisting of a two-stage ROI selection, a lightweight PCV streaming module, and an intelligent scheduler to balance the resources among communication memory and computation.
\textit{To address the first challenge}, we propose a two-stage efficient ROI selection method to dynamically focus transmission range, reducing network bandwidth, computational resources, and system overhead.
The first stage predicts the user's motion trajectory, combines the frustum cropping algorithm to select the initial ROI and defines the dynamic saliency intensity to further select related point cloud ROI.
The second stage proposes fine-grained ROI selection from the perspective of static saliency, including texture and geometric features, which reduces transmission data volume while ensuring PCV quality.
\textit{To address the second challenge}, we propose a lightweight encoder-decoder transmission network adapted to devices with different computing capabilities.
We analyze the existing GAN-based AI methods for point cloud reconstruction and propose a joint pruning and quantization approach to reduce the model size and parameters for deploying a decoder network on terminals.
Our lightweight encoder-decoder method leverages fine-tuning techniques in each pruning or quantization process to ensure accuracy.
Different pruning and quantization weights are defined for different devices to match the online codec AI model.
\textit{To address the third challenge}, we propose a DRL-based online intelligent scheduler considering dynamic network environments and computing capabilities of devices in AI-driven PCV streaming.
We design the reward of the DRL algorithm, including the transmission delay and the reconstructed quality, and describe matching DRL states, actions, and environments. 
We collect historical and online scheduling records to train the designed DRL model to optimize the intelligent scheduler iteratively.

We implement ISCom on mobile devices and conduct extensive experiments on three datasets, namely, S3DIS~\cite{armeni20163d}, 8iVFB~\cite{d20178i} and a Synthetic dataset.
We compare ISCom with baseline methods, including conventional video streaming techniques and AI-powered methods in video transmission quality, streaming efficiency, and runtime overhead.
Additionally, an in-depth analysis illustrates the designed modules of ISCom performing significant contributions to the improvements.
The results show that ISCom provides real-time PCV streaming services on mobile terminals between 10 and 22 FPS.
The key contributions of ISCom are fourfold:
\begin{itemize}[leftmargin=0.6cm,topsep=0.1cm]
	\setlength{\itemsep}{2pt}
	\setlength{\parsep}{0pt}
	\setlength{\parskip}{0pt}
	\item An interest-aware semantic communication scheme for PCV streaming on pervasive terminals.
	\item A two-stage point cloud ROI selection method based on user trajectory prediction, greatly improving the streaming efficiency and resource consumption.
	\item A lightweight AI transmission method adapted to various resource-constrained mobile terminals, solving the problems of high energy consumption and high computations.
	\item An intelligent AI codec model scheduling method based on DRL to enhance the adaptive characteristics of PCV transmission considering dynamic networks and uneven computing capabilities of terminals.
\end{itemize}

The remainder of the paper is organized as follows. We present the background and motivation in Section II and introduce the design of ISCom in Section III. Section IV presents the implementation, and Section V presents the evaluation in detail. Afterwards, we present related work in Section VI and we conclude our paper in Section VI.

\section{Background and Motivation}
We conduct a preliminary motivation study to illustrate the key challenges for achieving real-time PCV services. 
Specifically, we adopt three traditional compression methods, including Octree, Draco, MPEG G-PCC; and an AI-driven method, i.e., AITransfer. 
We transmit the down-sampled 8iVFB point cloud video~\cite{d20178i}, where each frame contains about 140,000 points. 
We use a server equipped with 48 Intel(R) Xeon(R) Gold 5118 CPUs @ 2.30 GHz for encoding and use a vivo IQOO smartphone equipped with a 2.92~GHz CPU for decoding. The transmission is done with a 5G network with a bandwidth of around 100~Mbps.
Figure~1 shows the required times for each phase in transmission.
\begin{figure}[htbp]
\centering
\centerline{\includegraphics[width=0.5\textwidth]{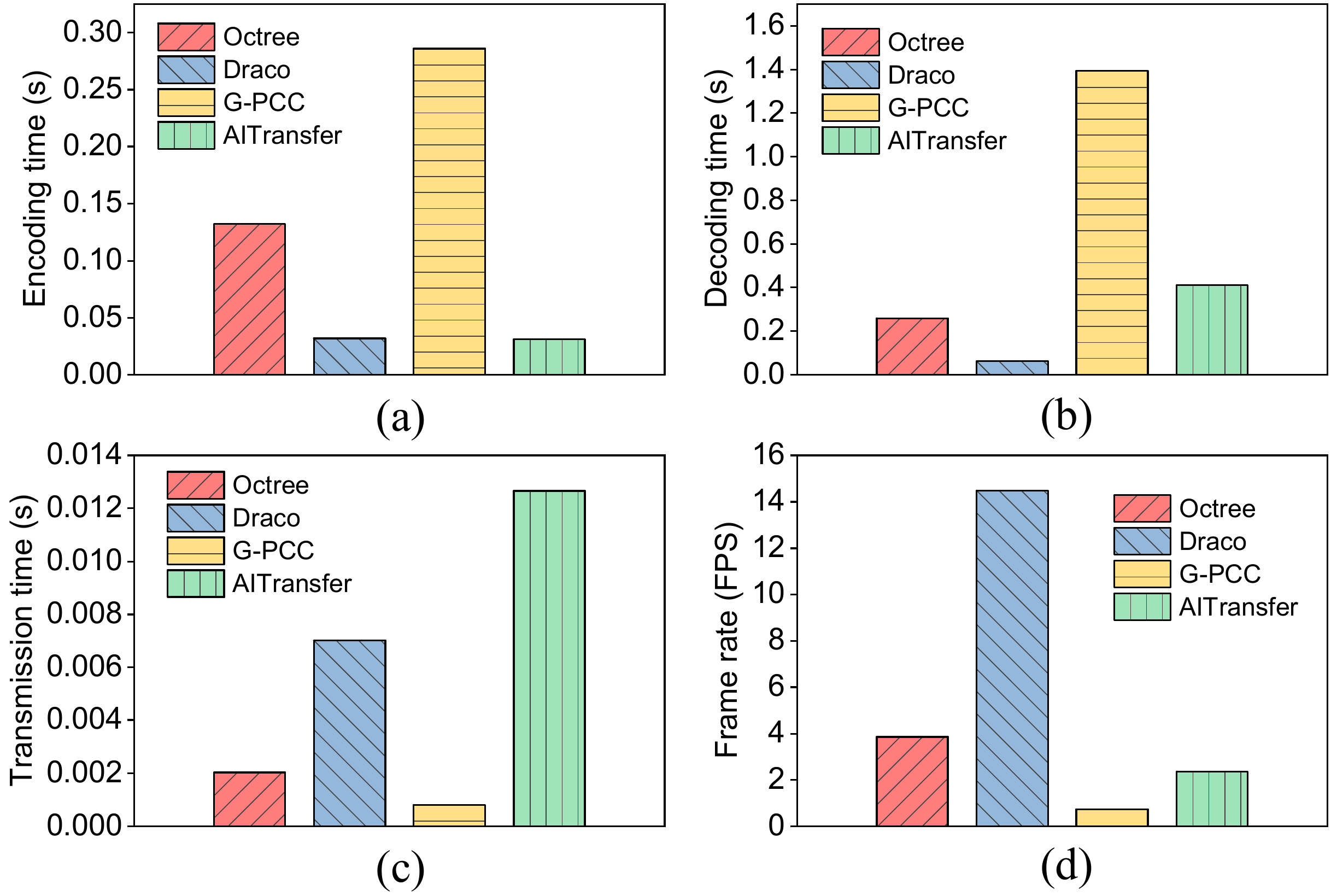}}
\caption{Streaming efficiency of baseline methods.}
\label{Fig_Motivation}
  	\vspace{-0.4cm}
\end{figure}

Our key observations are as follows: 1) all methods require a longer time decoding PCV than the encoding, which is because of the encoding on a powerful server;
2) 3D tree-based methods such as Octree and Draco have fast encoding and decoding rates, furthermore, due to a low compression rate, Draco achieves a faster encoding rate than Octree;
3) MPEG G-PCC has a high compression ratio, reflected by the transmission time, but fails to encode and decode PCV at a fast rate; 
and 4) although AI-native transmission methods, such as AITransfer, require deploying a neural model with a relatively small compression ratio to ensure the same quality, this results in long coding time and large data volume.

In summary, dominant methods can satisfy the bandwidth requirement by compression. 
Figure~1(d) shows that existing compression methods cannot stream the compressed PCV at 30~FPS even with a 5G network.
The main reason is that the high decoding cost on mobile devices limits the streaming efficiency, especially for streaming PCV with huge data volumes.
This motivates us to design an efficient transmission framework to enable an immersive PCV experience on mobile devices. 
On the one hand, we reduce the network bandwidth requirements for streaming PCV through a region of interest (ROI) selection module. 
On the other hand, we can also reduce the computation requirements on mobile devices by exploring a lightweight PCV encoder-decoder network.
For this new AI-native PCV delivery mechanism, it is clear that matching adaptive stream scheduling is also essential.

\begin{figure*}[!bp]
\vspace{-0.2cm}
 	\centering
 	\centerline{\includegraphics[width=0.95\textwidth]{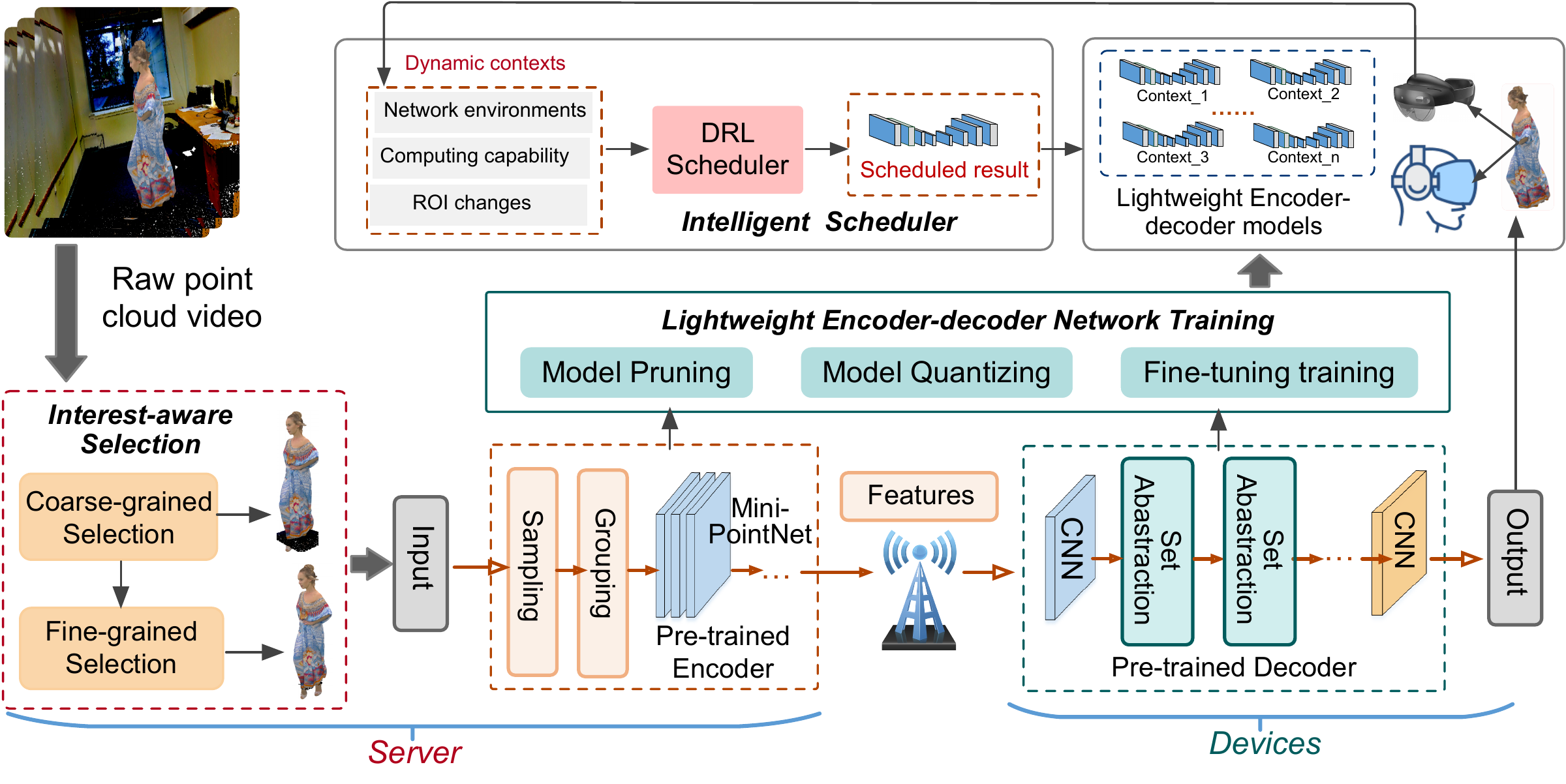}}
 	\caption{Overview of ISCom architecture.}
 	\label{Fig_1}
  	\vspace{-0.5cm}
 \end{figure*}
\section{Design of ISCom}
Figure~2 presents the design of ISCom, addressing how to provide a 6-DoF point cloud video streaming on the mobile device with three key modules, the interest-aware selection module, the lightweight semantic-aware transmission module, and an intelligent scheduler.

\subsection{Overview of ISCom}
ISCom leverages three key modules to provide stable real-time PCV transmission under dynamic transmission environments to various mobile devices. The interest-aware selection module extracts ROI regions from the original point cloud video to decrease data volume. 
Next, the lightweight encoder-decoder module provides efficient PCV transmission services on mobile devices. 
Besides, considering the fluctuation of the transmission environment, ISCom uses the intelligent DRL scheduler to provide stable PCV transmission quality.

\textbf{Interest-aware transmission model.}
This module is deployed on the edge server to reduce the size of the transmitted data. 
Through the two-stage ROI selection, the ROI in the PCV frame is extracted for later transmission. In detail, the complete point cloud video first goes through the coarse-grained ROI detection and selects the approximate key region information. Then, through fine-grained selection, the precise transmission region is determined. 

\textbf{Lightweight encoder-decoder transmission module.}
Through the lightweight AI-driven point cloud decoder design, ISCom can provide real-time PCV transmission services on most of the existing mobile devices. The processed point cloud frames are extracted with key features by the point cloud codec deployed on the edge server and sent to the mobile devices. The lightweight point cloud decoder will reconstruct the point cloud video on mobile devices when it receives the key features. 

\textbf{Intelligent scheduler.}
We deploy the intelligent DRL scheduler at the edge server to provide high-quality transmission services. It can sense fluctuation in the network bandwidth, the amount of transmitted data, and the computing power on a mobile device based on its feedback. By balancing the transmission accuracy and transmission delay simultaneously, the DRL-based scheduler can switch the optimal encoder-decoder transmission framework to ensure the quality of PCV transmission.

\subsection{Methods of ISCom}
\subsubsection{\textbf{Interest-aware transmission scheme}}
6-DoF point cloud video shares similar characteristics with 3-DoF video in that the user can only view a portion of the content within the current field of view, which is greatly influenced by the user's interests.~\cite{li2020joint}.
Therefore, accurate prediction of users' ROI can help us selectively reduce the amount of PCV transmission and dynamically provide on-demand streaming services.
\begin{figure*}[htbp]
	\centering
	\centerline{\includegraphics[width=0.95\textwidth]{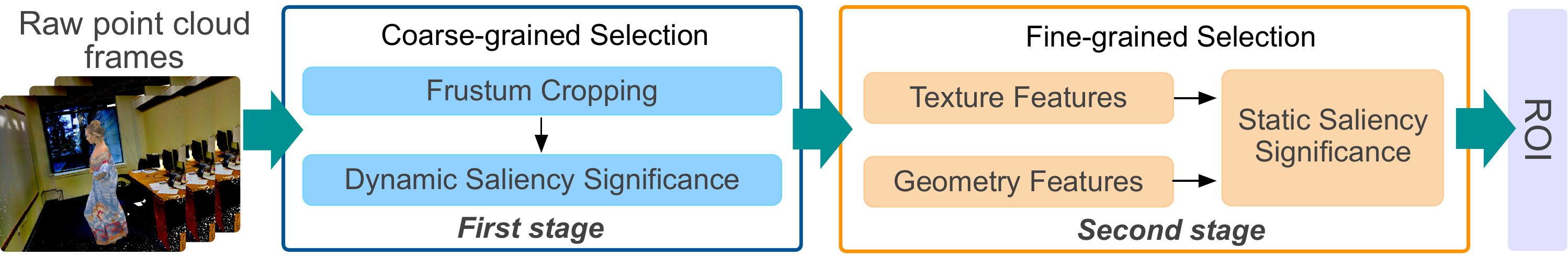}}
	\caption{Two-stage ROI selection.}
	\label{Fig_2}
	\vspace{-0.3cm}
\end{figure*}

As shown in Figure~3, we design a two-stage efficient ROI selection method for providing interest-aware PCV transmission in this module.
In the first stage, we propose a coarse-grained ROI selection, predicting users' trajectory from IMU sensors and employing the dynamic significance intensity to obtain a preliminary ROI point cloud.
Afterwards, the second stage is fine-grained ROI selection that calculates static saliency intensity of preliminary ROI point cloud and then supports accurate ROI selection.
Finally, we can downsample the point cloud regions according to the significance of the ROI so that the output point cloud information contains more points for the user-interest content and uses fewer points for non-ROI regions or content of lesser user interest.
Figure~4 presents an example of the two-stage ROI selection mechanism.
This module significantly reduces the data volume of transmission frames compared to the original frames.

\textbf{Coarse-grained ROI Selection.}
Since the user's field of view is limited, an accurate prediction of the user's field of view can be used to select the user's current ROI region for transmission.
We use the same model for user pose prediction as in~\cite{hou2019head}, using the previous $k$ frames of the IMU data as input and outputting the estimated pose for the next $k$ frames.
\begin{equation}
	\setlength{\abovedisplayskip}{0pt}
	\setlength{\belowdisplayskip}{0pt}
	Pose(k)= \mathcal{F}(p_t, p_{t-1},..., p_{t-k}),
\end{equation}
where $Pose(k)=[p_{t+1},p_{t+2},... ,p_{t+k}]$ denotes the IMU data in the next $k$ frames.
Then we can calculate the user motion trajectory based on the estimated pose.
Assuming that the user is located at the virtual camera in Figure~4, we can obtain the coarse-grained viewing content based on the perspective projection matrix and frustum cropping algorithm~\cite{assarsson}.
For the point $v$ in the original PCV frame $F$, a coarse-grained ROI set $\hat{F}$ can be derived after frustum cropping.
 \begin{figure*}[htbp]
	\centering
	\centerline{\includegraphics[width=0.95\textwidth]{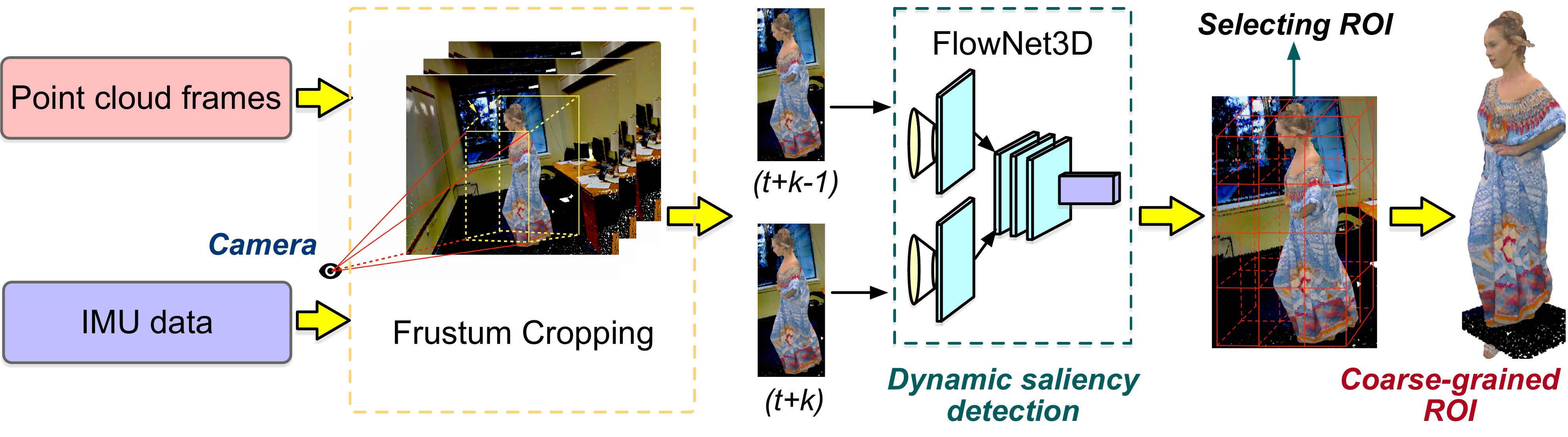}}
	\caption{Coarse-grained ROI selection.}
	\label{Fig_3}
	\vspace{-0.3cm}
\end{figure*}

Next, we introduce dynamic saliency detection for a precise selection considering the motion features.
Dynamic saliency detection is crucial due to the fact that the attentions and interests of users usually focus on the moving contents.
We use the FlowNet3D~\cite{liu2019flownet3d} network to extract the motion features and define the dynamic saliency significance of the ROI to select the ROI regions.
Specifically, we input $\hat{F}_{t-1}$ and $\hat{F}_t$ to FlowNet3D for prediction and output the motion components $\vec{e_i}$ of each point $v'=(x',y',z')$.
Then, we divide $\hat{F}$ into $T_1$ blocks and calculate the average movement $S_E^D$ of all points in each block according to the dynamic significance intensity as follows:
\begin{equation}
	\setlength{\abovedisplayskip}{4pt}
	\setlength{\belowdisplayskip}{3pt}
	S_E^D=\frac{\sum_{i=1}^{N}}{||e_i||},
\end{equation}
where $e_i$ is the flow field intensity of each point $p_i$, indicating the movement range.
Finally, we select a 60\% ROI based on the average flow field intensity as the final result of coarse-gained ROI selection.

\textbf{Fine-grained ROI Selection.}
In the next step, we propose to select a more accurate ROI from the static fine-grained saliency.
Compared with dynamic saliency features, static saliency focuses on the geometric and textural features, which also considers the user's viewing scale. 
First, we divide the output ROI from the first stage into $T_2$ dense blocks of uniform size $L*L*L$.
We define the effect of user viewpoint on static saliency as the viewpoint descriptor from each block centre $o_i(x_i,y_i,z_i)$ to the user viewpoint $v(x_c,y_c,z_c)$, as follows:
\begin{equation}
	\setlength{\abovedisplayskip}{4pt}
	\setlength{\belowdisplayskip}{3pt}
	S_v^D=\frac{\beta}{\ln (\phi)}+(1-\beta)\cos (\theta),
\end{equation}
where $\beta$ is the weight of balancing distance and angle, $\phi=||\overrightarrow{o_i v}||$ is the distance from the centre position $o_i$ of each block to the user's viewpoint $v$.
A larger $\phi$ means the farther the user's viewpoint is from the block, the less significant it is.
$\theta$ denotes the angle between $\phi$ and the user viewpoint angle toward $w$ as follows:
\begin{equation}
	\setlength{\abovedisplayskip}{4pt}
	\setlength{\belowdisplayskip}{3pt}
	\theta = \frac{\overrightarrow{o_i v}\cdot w}{|\overrightarrow{o_i v|}\cdot{|w|}}.
\end{equation}

Note that the larger the $\theta$ is, the lower the static significance.
Next, we construct the geometric texture descriptor~\cite{li2022optimal} to represent the chunk geometry features and texture differences, as follows:
\begin{equation}
	\setlength{\abovedisplayskip}{4pt}
	\setlength{\belowdisplayskip}{3pt}
	T_i^E = 1- \exp (-\frac{1}{R}\sum_{j=1}^{R}\frac{\psi^2(t_i,t_j)}{1+|t_i - t_j|}),
\end{equation}
where $\psi^2(t_i,t_j)$ is the geometric texture feature distance between two points, which can be expressed as:
\begin{equation}
	\setlength{\abovedisplayskip}{4pt}
	\setlength{\belowdisplayskip}{3pt}
	\psi^2(t_i,t_j) = \chi^2(t_i,t_j)+\lambda \cdot \gamma^2(t_i,t_j).
\end{equation}

$\lambda$ is a weighting factor to balance the weight of geometrics and texture.
We set the same $\lambda$ as 0.35 as the~\cite{li2022optimal}.
Based on the above definition, we can calculate the salient static features of ROI as follows:
\begin{equation}
	\setlength{\abovedisplayskip}{0pt}
	\setlength{\belowdisplayskip}{0pt}
	S^E= S_v^E\cdot S_i^E.
\end{equation}

Finally, we downsample the coarse-grained ROI according to the static salient, which means that a higher downsampling rate is suitable for low static saliency ROI.
In addition, we show an example in Figure 4, using the interest-aware module to obtain the ROI point cloud and the edges perform smoothly by downsampling with the static saliency.

\subsubsection{\textbf{Lightweight PCV encoder-decoder network}}
Interest-aware selection effectively reduces the point cloud transmission scale.
However, the user's ROI still contains many dense points (e.g., longdress's tasks still contain 800,000 points).
Therefore, it is difficult to provide real-time PCV transmission for resource-constrained devices.
The existing AI-drive encoder and decoder in AITransfer~\cite{huang2021aitransfer} can significantly reduce transmission data volume.
However, this AI-driven method employs a GAN-based generative network for reconstruction, requiring heavy computational resources.
In addition, AITransfer's encoder and decoder have large parameters and require long training time and high inference latency, which obviously cannot provide real-time PCV streaming service on devices.

\textbf{Pruning process.}
To this end, we propose a lightweight encoder-decoder network approach for resource-constrained devices.
The approach jointly prunes and quantises the pre-trained model to reduce the parameters and model size for accelerating inference.
Specifically, we use AITransfer~\cite{huang2021aitransfer} as a pre-trained model for training our lightweight encoder-decoder network.
This process includes model pruning, quantisation, and fine-tuning processes.
Given a pre-trained full-precision encoder-decoder model, it has $N$ convolutional neural network~(CNN) layers and $M$ fully connected layers.
We define $W_{i}^{(n)}$ as the weight of the $i^{\rm th}$ channel in the $n^{\rm th}$ CNN layer.
Our pruning process uses the same loss function of AITransfer as follows:
 \begin{equation}
 	\setlength{\abovedisplayskip}{4pt}
 	\setlength{\belowdisplayskip}{3pt}
 	\mathcal{L}=\lambda_{rec}\mathcal{L}_{rec}+||\theta||^2,
 \end{equation}
where $\lambda_{rec}$ balances the EMD loss $\mathcal{L}_{rec}$ and the loss on the point cloud rotation $||theta||^2$.
 \begin{equation}
	\setlength{\abovedisplayskip}{4pt}
	\setlength{\belowdisplayskip}{3pt}
	\mathcal{L}_{rec} = \min_{\phi:P\leftarrow Q}\sum_{p_i \in P}||p_i - \psi(p_i)||_2,
\end{equation}
where $P$ and $Q$ denote the reconstructed point cloud and the original point cloud, respectively.
During pruning, if the current loss is below the set threshold $l_{th}$ for triggering the pruning operation, 
we can calculate the $W_{th}^{(k)}$ according to the pruning weight ratio $\zeta$ for each round as follows:
\begin{equation}
	\setlength{\abovedisplayskip}{4pt}
	\setlength{\belowdisplayskip}{3pt}
	W_{th}^{(k)}=\left\{
	\begin{array}{lr}
		S_{C\times \zeta}, k \in (1,N),\\
		\vspace{-0.2cm}
		\\ S_{M\times \zeta}, k \in (1,M). \\ 
	\end{array}
	\right.
\end{equation}

$C$ is the number of CNN layer channels, and $M$ is the total number of fully-connected layer weights.
$S$ represents ranking each layer's weights according to its importance.
The remaining parameters can be pruned layer by layer using the following equation:
\begin{equation}
	\setlength{\abovedisplayskip}{4pt}
	\setlength{\belowdisplayskip}{3pt}
	W_{i}^{(n)}=\left\{
	\begin{array}{lr}
		W_{i}^{(n)}, |W_{i}^{(n)}| > 0\\
		\vspace{-0.2cm}
		\\ 0, otherwise \\ 
	\end{array}
	\right.
\end{equation}

After each pruning step is completed, the gradient is calculated for the remaining parameters according to the loss function.
When completing the setting epoch $E$, we repeat the process mentioned above until we reach $\zeta$ or end the pruning when completing the setting epoch $E$.

\textbf{Quantization process.}
We use the following formula to quantise the weights and the inputs from 32-bit to 8/16-bit to improve the inference of the decoder.
The process of quantifying the weights is as follows:
 \begin{equation}
	\setlength{\abovedisplayskip}{4pt}
	\setlength{\belowdisplayskip}{3pt}
	\hat{W}_{i}^{(n)}=round(q_w(W_{i}^{(n)}-min(W^{(n)}))),
\end{equation}   
where $q_w$ is the scaling factor that maps the weights from 32-bits float to integer, defined as:
 \begin{equation}
	\setlength{\abovedisplayskip}{4pt}
	\setlength{\belowdisplayskip}{3pt}
    q_w = \frac{2^m-1}{max(W^{(N)})-min(W^{(N)})},
\end{equation}   

For the input quantization, the selection of the quantification range interval significantly impacts the accuracy. 
Therefore, we remove significant outliers before performing quantization.
Algorithm~1 presents a pseudo-code description for training the lightweight encoder-decoder network.
\IncMargin{0.2em}
\begin{algorithm}[!htb]
	\small
	\setlength{\abovedisplayskip}{-3pt}
	\setlength{\belowdisplayskip}{-10pt}
	\caption{Lightweight encoder-decoder network training}
	\label{alg::CoScheduler}
	\KwIn{$model$: pre-trained model, $\zeta$: pruning sparse ratio  }
	\KwOut{$model'$: jointly pruned and quantised model.}
	
	\FunctionPruning{\textbf{Return} $W$}{
		$C \Leftarrow WeightNumberCount(W)$\;
		$S\Leftarrow Sort(W)$\;
		\ForEach{$w\in W$}{
			\tcc{$\zeta$ is the pruning sparsity}
			\If{$w< S_{c\times \zeta}$}{
			$w\Leftarrow 0$	
				\vspace{-0.05cm}
		}
		\vspace{-0.05cm}	
		}
	}
	\FunctionQuantizing{\textbf{Return} $W$}{
		$W_{max}\Leftarrow \max(W)$\;
		$W_{min}\Leftarrow \min(W)$\;
		$q_w\Leftarrow Eq.(13)$\;
		\ForEach{$w \in W$}{
			$w\Leftarrow Eq.(12)$
		}
	}
	\FunctionProcedure{\textbf{Return} $model'$}{
		$N\Leftarrow CountModelLayer(model)$\;
		\For{$n\in N$}{
			$W^{n}\Leftarrow \textsc{Pruning} (W^{n})$\;
		}
		$model' \Leftarrow FineTuneTraining(model)$\;
		\For{$n\in N$}{
			$W^{n}\Leftarrow \textsc{Quantizing} (W^{n})$\;
		}
	}
		\vspace{-0.05cm}
\end{algorithm}
\DecMargin{0.2em}

\subsubsection{\textbf{Intelligent Scheduler}}
Since network environments and the computing capability of devices affect the PCV streaming, it requires an intelligent scheduler that considers both factors to provide the best lightweight encoder-decoder model for dynamic transmission environments.
In this section, we design a dynamic transmission scheduling method that senses contexts by collecting the network bandwidth and state of devices.
Then, it can adjust the codec model for optimal streaming experience in such dynamic environments based on DRL.

\begin{figure}[htbp]
	\centering
	\centerline{\includegraphics[width=0.45\textwidth]{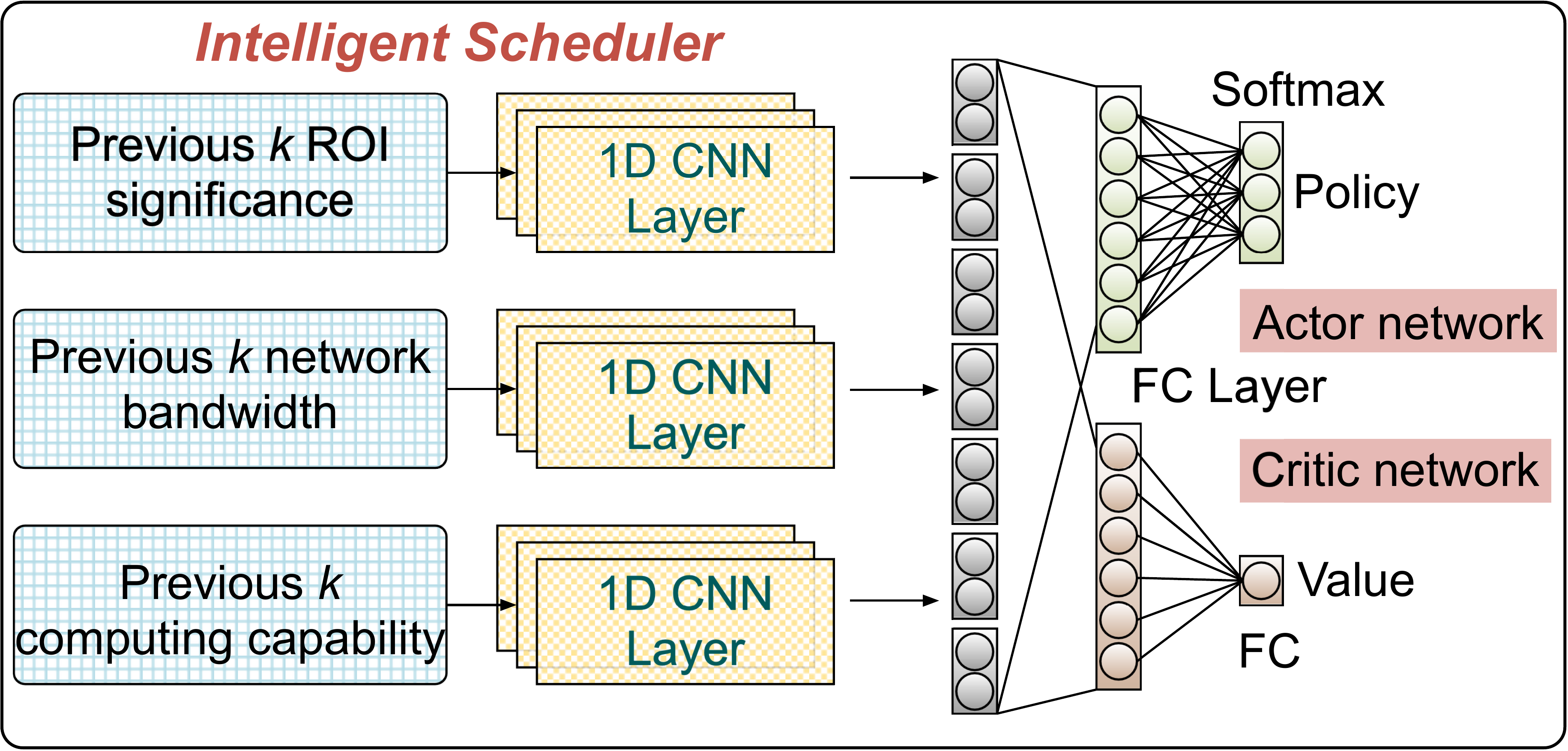}}
	\caption{The NN architecture of intelligent scheduler.}
	\label{Fig_4}
\end{figure}

\textbf{State design.}
The state is a set of variables describing the state of the environment $S_t$ at the current time $t$.
In PCV streaming, the state involves the user ROI, the computing capability of the terminal, and network bandwidth.
Hence, we define the state as $S_t=(\vec{n_t},\vec{c_t},\vec{b_t})$, where $\vec{n_t}=(n_{t-k},n_{t-k+1},...,n_{t-1})$ denotes the ROI significance value in the previous $k$ frames, reflecting the fluctuation of data volume per frame.
$\vec{c_t}=(c_{t-k},c_{t-k+1},...,c_{t-1}$ is the estimated required computing capability or previous $k$ frames, representing the changes of mobile devices, which can be estimated from the reconstruction delay in the previous $k$ frames.
$\vec{b_t}=(b_{t-k},b_{t-k+1},...,b_{t-1})$ is the average network bandwidth in previous $k$ frames, reflecting the bandwidth fluctuations.

\textbf{Action and Policy.}
We introduce the action space by selecting various trained lightweight encoder-decoder transmission models for state $S_t$.
Hence, the policy is the probability distribution of mapping from state space to action space with the highest probability.
As shown in Figure~5, we use a fully concatenated layer to process all the input and extract the underlying features between the inputs.
Besides, the actor-network uses another fully concatenated layer to process the relationship between features and finally feeds the inputs into the Softmax layer to obtain the probability distribution.
Then, we train the parameters in the policy network to get the optimal scheduling policy.

\textbf{Reward.}
We propose a reward function applicable to the PCV transmission as follows:
 \begin{equation}
	\setlength{\abovedisplayskip}{4pt}
	\setlength{\belowdisplayskip}{3pt}
	r_t = \eta \cdot f_t +(1-\eta)L_{P_i},
\end{equation}
where $f_t$ is the frame rate at $t$ moments, $L_{P_i}$ is the transmission accuracy of the selected $P_i$, and $\eta$ is the weight to balance the transmission quality.

For the reward design, the transmission performance of image or video streaming is measured by bit rate-distortion optimization (RDO).
The principle is to optimize the compression ratio and reconstruction of video quality.
Consequently, we design the reward considering these two aspects into consideration. 
One is to represent the distortion through the accuracy $L_{P_i}$, and the other is to represent the bit rate through the frame rate $f_t$.
To sum up, we use a weighted sum of the two objects for training simplicity.

\textbf{Training DRL for intelligent scheduling.}
We train the intelligent scheduler by using A3C algorithms~\cite{mnih2016asynchronous} because of its popular characteristic~\cite{mao2017neural}.
Specifically, A3C has a global network and multiple workers. Each worker updates the network parameters it has learned from the task to the global network.
Then, each worker pulls the global parameters in the next learning.
A3C maximizes the expected cumulative reward by training two types of neural networks, i.e., actor-network and critic network.
The actor-network inputs policy $\pi$, and the critic network helps to train the actor-network.

We train the actor-network by the policy gradient method~\cite{mnih2016asynchronous}. 
The cumulative gradient with respect to the policy parameters $\theta$ is calculated as:
\begin{equation}
d\theta \gets d\theta + \bigtriangledown _{\theta'}\log_{}{\pi (a_{i}|s_{i}; \theta')(R-V(s_{i};\theta'_{v}))} ,
\end{equation}
where $R$ is the cumulative discounted reward, and $V(\cdot)$ is the state value function.
The cumulative gradient with respect to the critic parameters $\theta_{v}$ is calculated as:
\begin{equation}
d\theta_{v} \gets d\theta_{v} + \partial  (R-V(s_{i};\theta'_{v}))^{2} /\partial \theta_{v}.
\end{equation}

A3C uses an entropy regularization term for the actor’s update helping the network converge to a better policy~\cite{mnih2016asynchronous}.

\section{Evaluation}
\subsection{Implementation}
We deploy the ISCom system flow (Figure 1) on high-performance edge servers and Android smartphones. 
For the interest-aware selection module, we modify the source code of Flownet3D using TensorFlow and the OpenGL and Open3D libraries to implement the two-stage ROI selection. 
In the lightweight PCV codec module, we optimize the source code of AITransfer using TensorFlow, then prune and quantise the pre-trained model, and then convert the model to the TensorFlow-lite format for future reuse. 
For mobile devices, we use the TensorFlow Lite Interpreter AIP to deploy the lightweight point cloud decoder into a Java Android application and use the Java OpenGL library to render point cloud video on mobile devices. 
For the intelligent scheduler, the deep learning agent is implemented using the TensorFlow framework, with hyper-parameters tuned empirically.

\subsection{Experiment setup}
\subsubsection{\textbf{Datasets}}
We conduct extensive experiments using three types of point cloud datasets, including Stanford 3D Indoor Scene (S3DIS)~\cite{armeni20163d}, 8i Voxelized Full Bodies (8iVFB)~\cite{d20178i}, and a synthetic dataset.
\begin{itemize}
    \item \textbf{S3DIS.} 
    The S3DIS dataset~\cite{armeni20163d} comprises of 6 large-scale indoor areas with 271 rooms. 
    Each room contains 1.1 to 1.5 million points, and each point is annotated with one of 13 categories. 
    We select the office area to verify the semantic understanding ability of ISCom on the whole point cloud scene.
    \item \textbf{8iVFB.}
    The 8iVFB dataset~\cite{d20178i} provides four dynamic point cloud sequences: Longdress, Loot, Redandblack, and Soldier.
    Each sequence records a moving human subject at 30~FPS, over a 10~s period, where each frame contains 0.8 to 1.1 million points.
    We select this dataset to verify the performance of the transmission network in ISCom. 
    \item \textbf{Synthetic Dataset.}
    Since there is no open source full scene PCV benchmark dataset, we manually synthesize a full scene dataset containing background and motion subjects to verify the effectiveness of the interest-aware transmission scheme.
    The synthetic dataset comprises of five rooms in the S3DIS Office area and 100 consecutive frames of the Longdress subject in 8iVFB, for a total of $5\times100=500$ frames.
    Specifically, we start with normalizing S3DIS and 8iVFB frames to adjust each object to the same scale and align the origin of coordinates. 
    Then, we downsample the Longdress subject to 0.2 million points and make it match the Office area by scaling rotation and translation operations.
    Figure~6 presents five representative synthetic scenes in the synthetic dataset. 
\end{itemize}

\begin{figure}[!htb]
	\centering
	\subfigure[Scene I]{\includegraphics[width=0.15\textwidth]{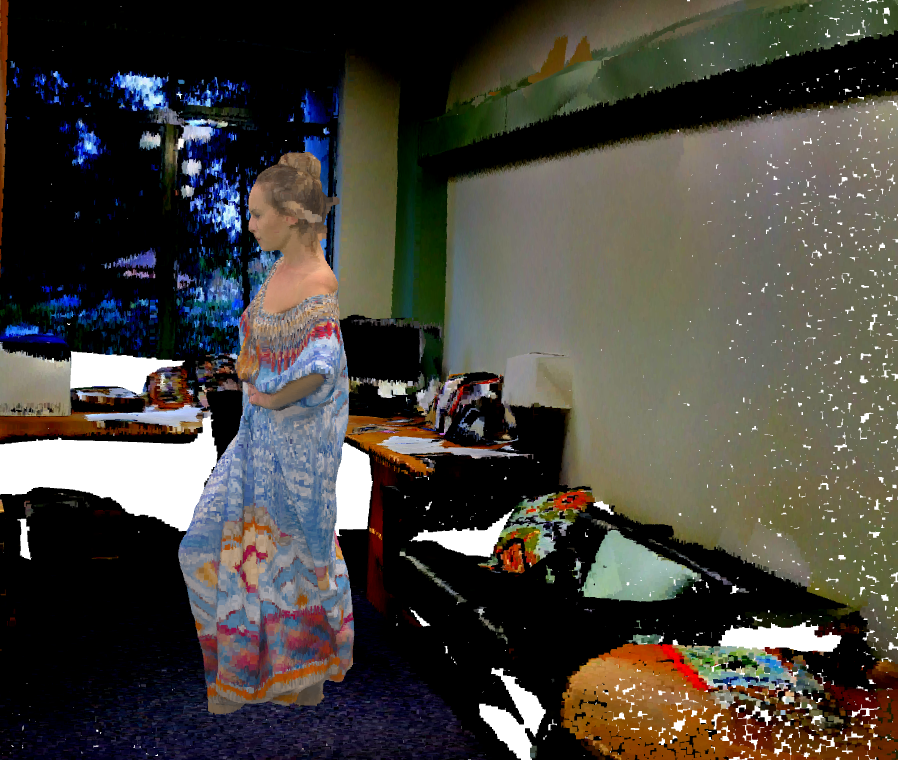}}
	\subfigure[Scene II]{\includegraphics[width=0.15\textwidth]{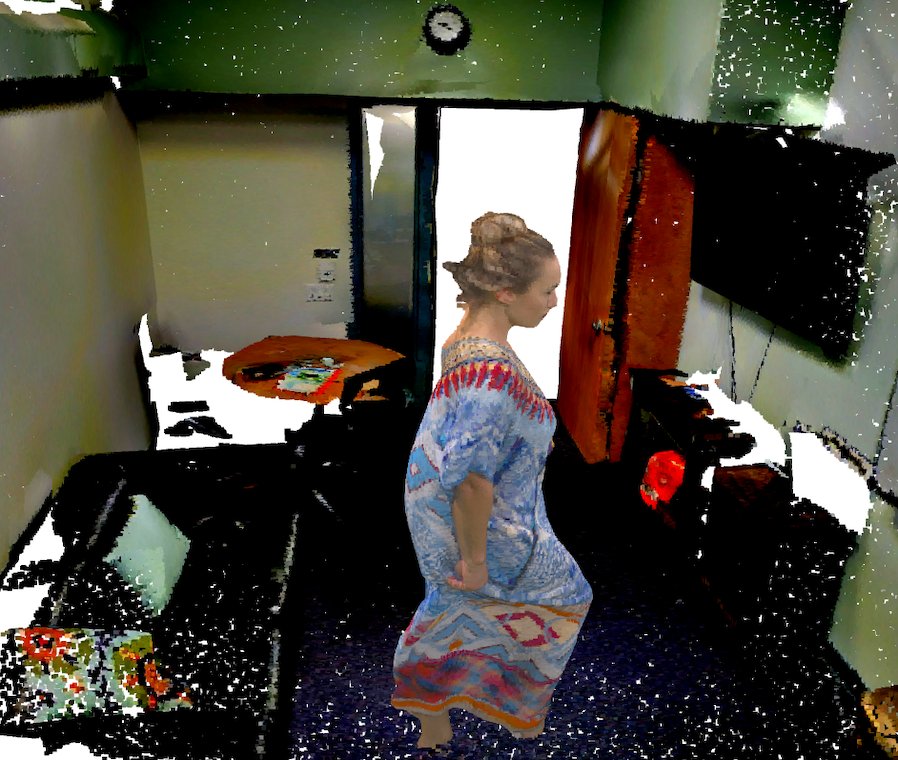}}
	\subfigure[Scene III]{\includegraphics[width=0.15\textwidth]{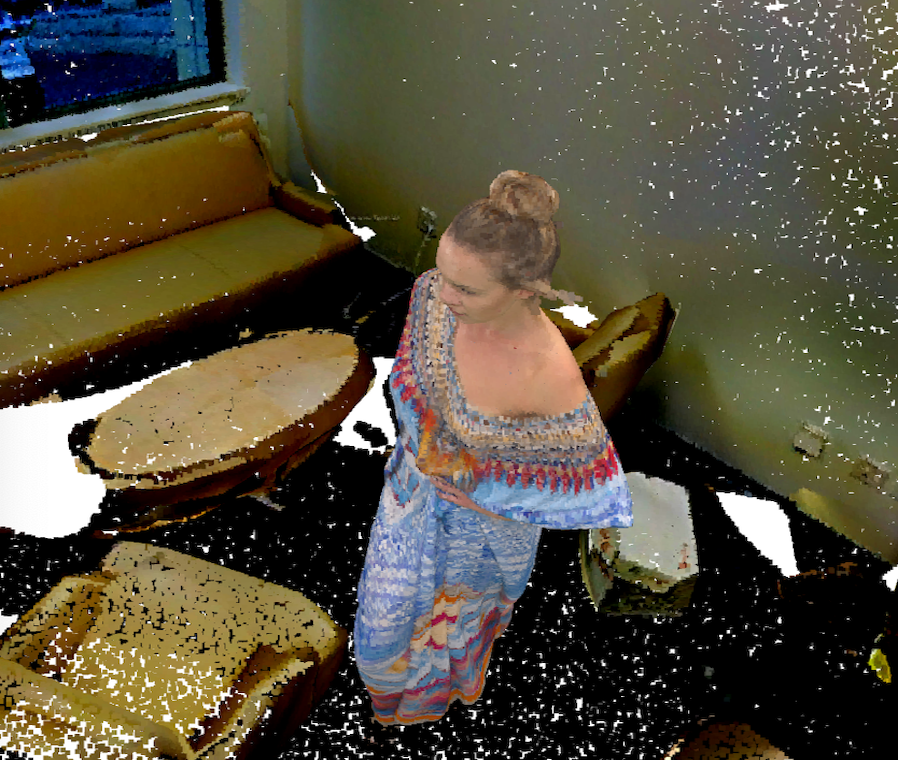}}
	\subfigure[Scene IV]{\includegraphics[width=0.15\textwidth]{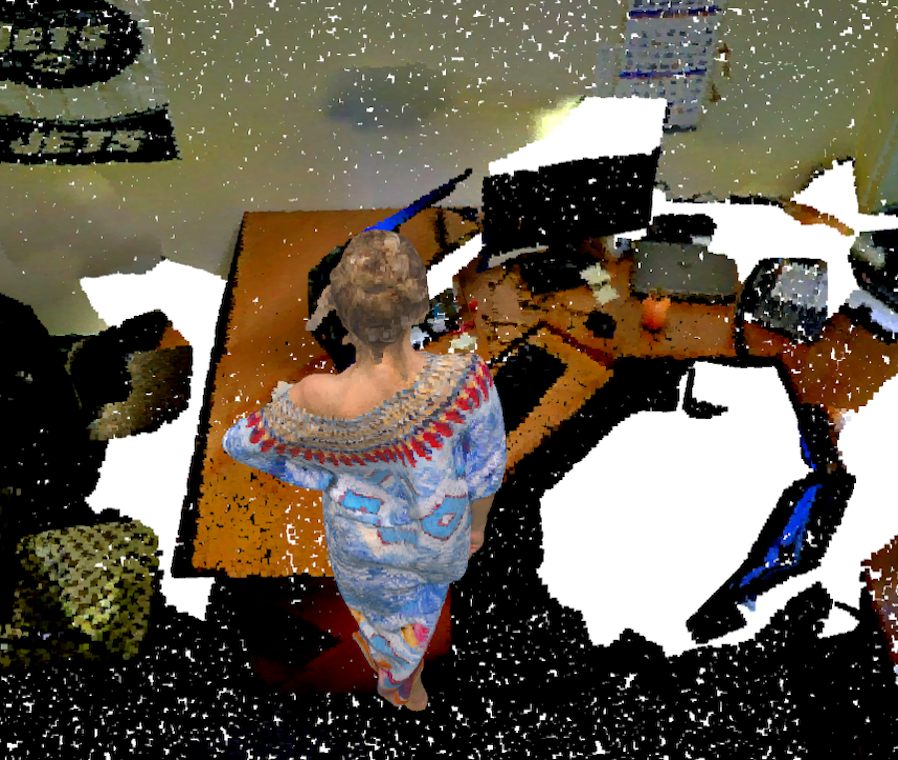}}
	\hspace{0.15cm}
	\subfigure[Scene V]{\includegraphics[width=0.15\textwidth]{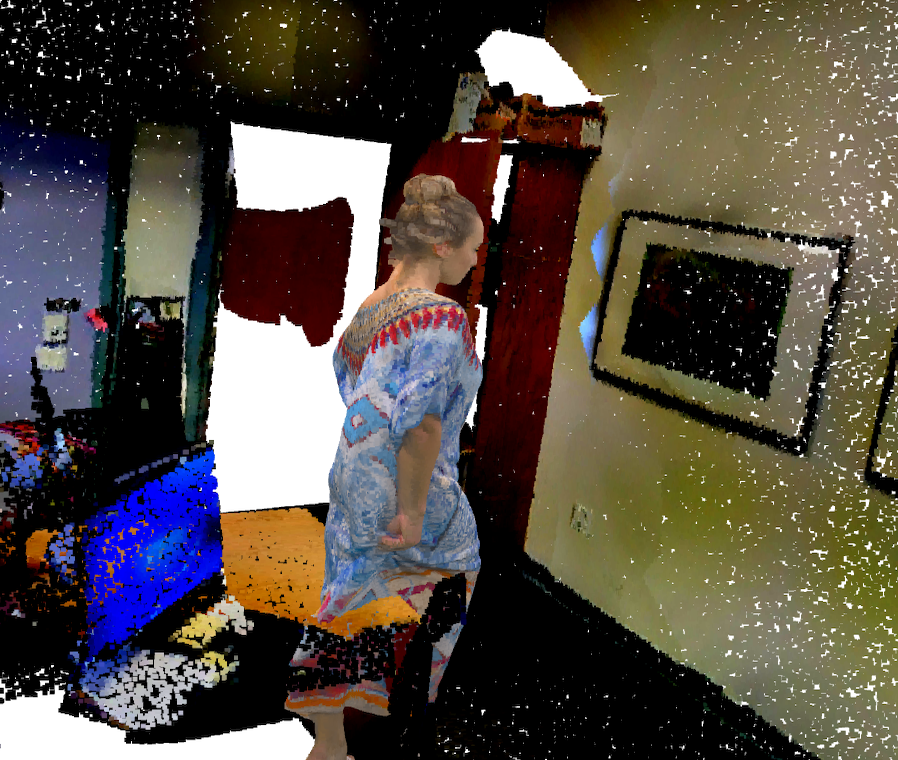}}
	\caption{Synthetic datasets by placing with Longdress subject in various scenes.}
	\label{Fig_6}
	\vspace{-0.5cm}
\end{figure}

\subsubsection{\textbf{Baselines and Methods}}
We evaluate ISCom with three traditional compression-based systems (i.e., Octree~\cite{schnabel2006octree}, Draco~\cite{draco}, and MPEG G-PCC~\cite{graziosi2020overview}); and an AI-driven transmission system (i.e., AITransfer~\cite{huang2021aitransfer}).

\paragraph{\textit{\textbf{Traditional compression-based systems}}} 
Most of recent systems deliver point cloud video using basic compression techniques. To evaluate ISCom under different network and terminal environments, we have implemented three classical compression method-based transmission systems, as follows:
\begin{itemize}
    \item \textbf{Base-Octree.}
    The principle of Octree~\cite{schnabel2006octree} is recursively dividing a cubical volume of space into eight equivalent of sub-cubes.
    Varying the depth can achieve different compression ratios.
    Octree is commonly used in Point Cloud Library (PCL)~\cite{Rusu_ICRA2011_PCL} for point cloud compression.
    \item \textbf{Base-Draco.}
    Google's Draco~\cite{draco} is an efficient 3D data compression method. 
    Different from Octree, Draco adopts kd-tree structure for point cloud representation and partition.
    Draco can also achieve different compression ratios by adjusting the depth parameter.
    \item \textbf{Base-G-PCC.}
    MPEG standard provides a point cloud codec with powerful compression capability exceeding that of current approaches.
    For experimental fairness, we select the Geometry-based Point Cloud Compression (G-PCC)~\cite{graziosi2020overview}, which independently encodes the point cloud frames in the same setting as other baselines. 
\end{itemize}

\paragraph{\textit{\textbf{AI-driven transmission system}}}
AITransfer~\cite{huang2021aitransfer} is a representative complete AI-driven PCV streaming system that designs an end-to-end neural network to extract key semantic features from the point cloud and recover the features back to the original video.
The high-level idea of AITransfer is to deliver key semantic information rather than the original point cloud or compressed data, reducing the data volume and bandwidth consumption. 

\subsubsection{\textbf{Evaluation Metrics}}
We evaluate ISCom with baselines from three aspects, i.e., video quality, streaming efficiency, and system overhead. 
\begin{itemize}
    \item \textbf{Video quality}.
    The video quality affects the users' experience straightforwardly.
    The distortion caused by compression and transmission is an important metric to measure the video streaming system.
    This means that the reconstructed video should be consistent with the original video.
    Peak Signal to Noise Ratio (PSNR) and Structural Similarity (SSIM)~\cite{hore2010image} are two widely used metrics in image and video fields.
    However, considering the fairness of experiments, we transmit point cloud frames without colour information as the same as AITransfer, Therefore, these two metrics are not usable.
    In this work, we adopt two commonly used geometry-based metrics to measure the quality of point cloud frame quality, including Chamfer Distance (CD)~\cite{butt1998optimum} and Hausdorff Distance (HD)~\cite{berger2013benchmark}. 
    The greater these distances are, the lower the accuracy of the reconstructed video is.
    \item \textbf{Streaming efficiency}.
    Streaming efficiency is another important metric to measure the played video quality.
    The efficiency involves three phases, including encoding time, decoding time, and transmission time. 
    These three types of time consumption measure the encoding capacity, decoding capacity, and compression ratio, respectively, which jointly determine the maximum frame rate of the played video.
    \item \textbf{System overhead}.
    We measure the system overhead of different comparison methods from four aspects, i.e., CPU usage, memory usage, and device temperature.
    We record these metrics using the commercial software PerfDog~\cite{perfdog} when conducting the video decoding task on different devices. 
    To ensure the fairness of the experiment, devices only run experimental applications and the screen brightness is set to the lowest.
\end{itemize}

\subsection{Performance of video streaming}
\subsubsection{\textbf{Evaluation of video quality}}
\begin{figure}[!htb]
	\begin{minipage}[t]{\linewidth}
		\centering
		\subfigure[Peformance of \textit{device-1} in 5G]{
			\includegraphics[width=4.25cm]{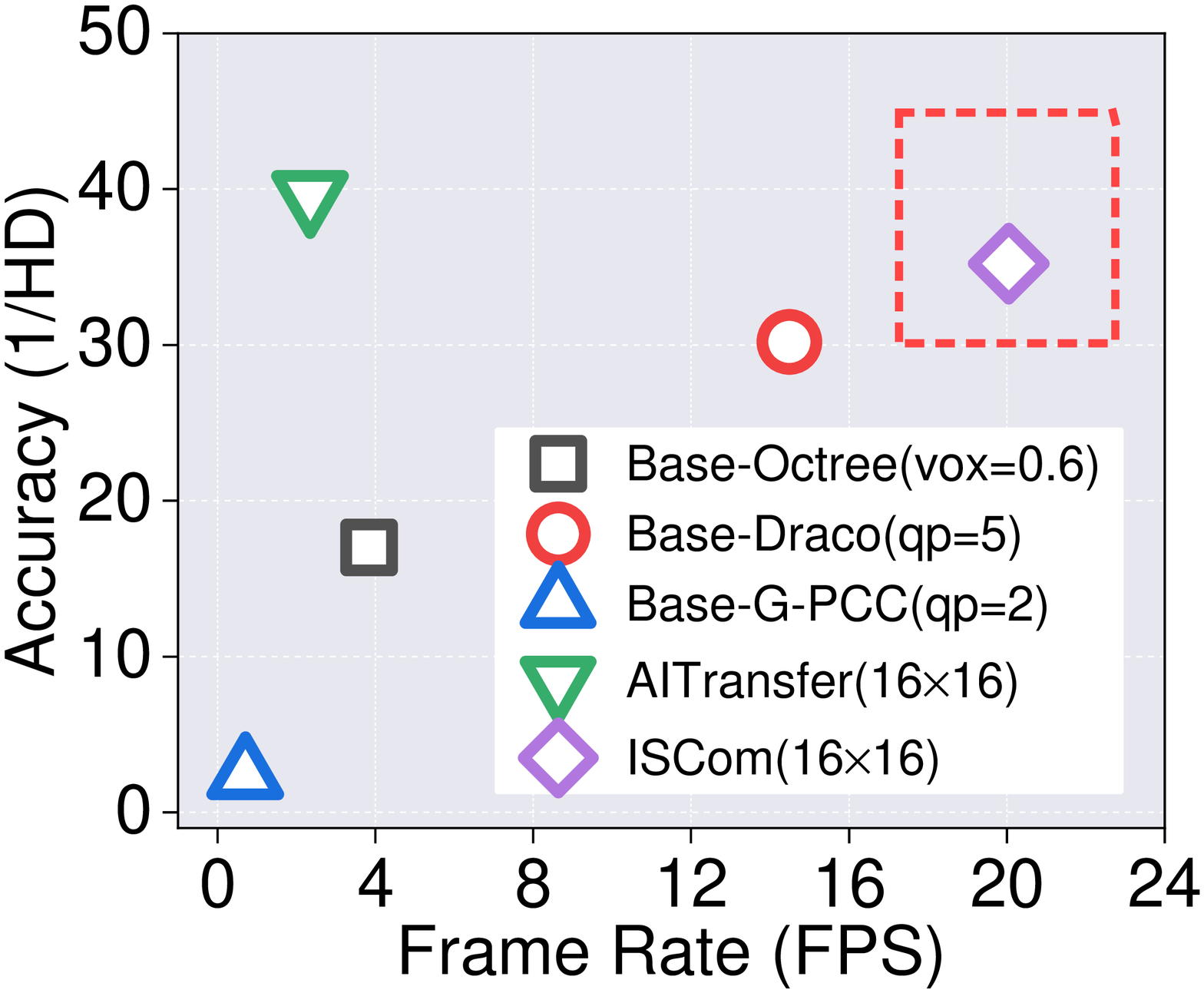}}
		\subfigure[Peformance of \textit{device-3} in 3G]{
			\includegraphics[width=4.25cm]{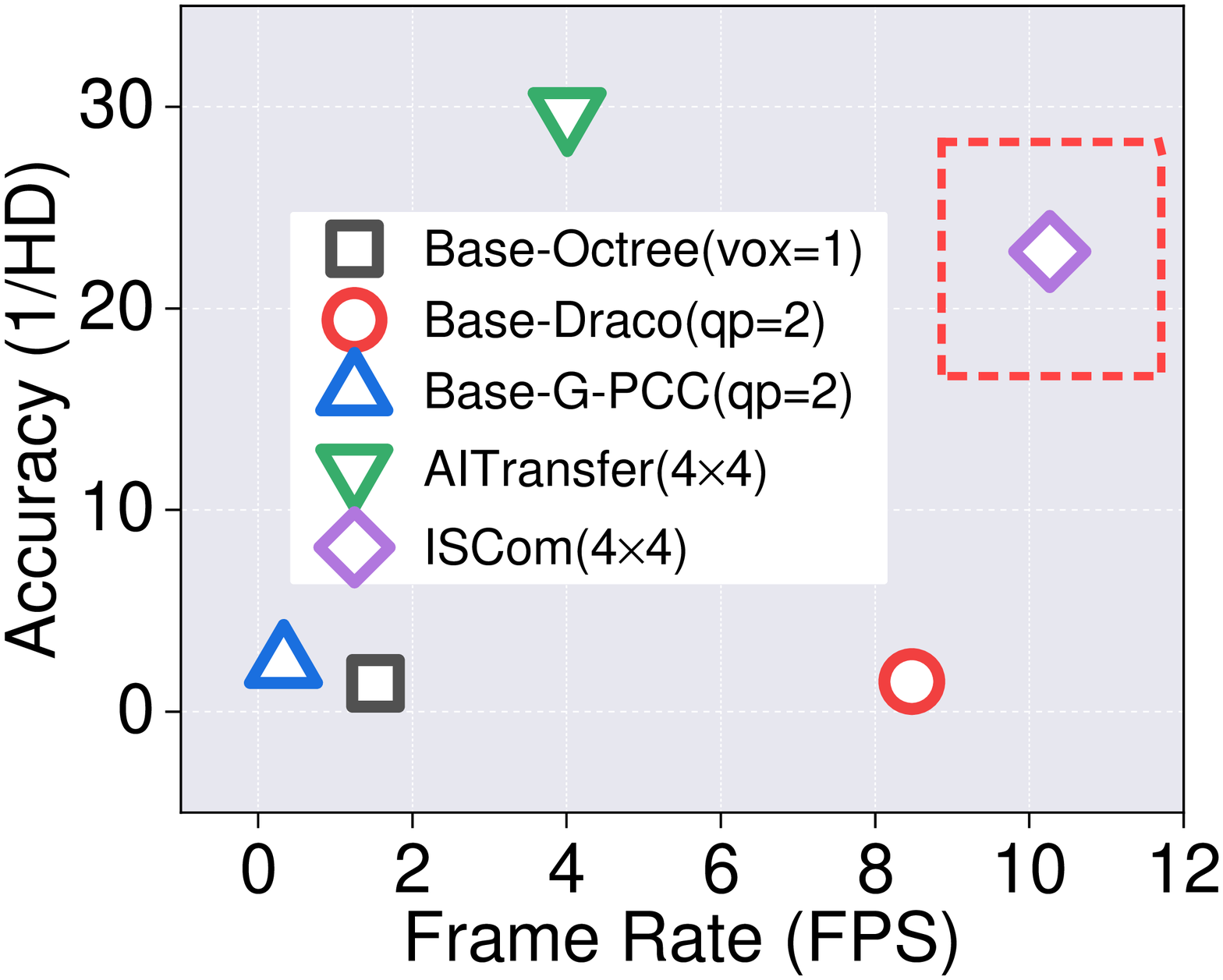}}
		\caption{Results of accuracy (1/HD) and frame rate}
		\label{Fig521HD}
	\end{minipage}%
	\vspace{-0.2cm}
\end{figure}

\begin{figure*}[!htb]
	\begin{minipage}[t]{\linewidth}
		\centering
		\subfigure[Accuracy and FPS in 3G]{
			\includegraphics[width=4.3cm]{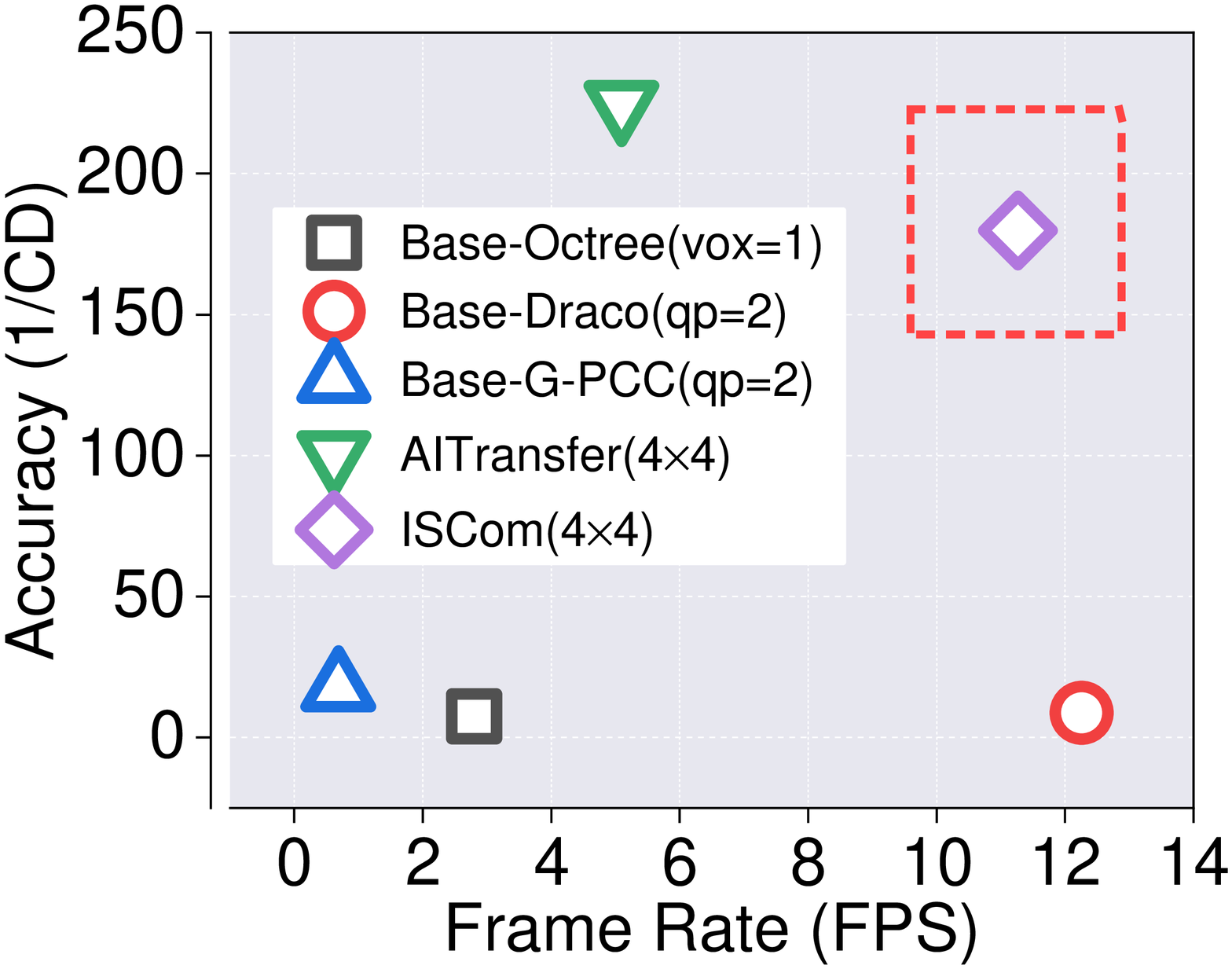}}
		\subfigure[Accuracy and FPS in 4G]{
			\includegraphics[width=4.3cm]{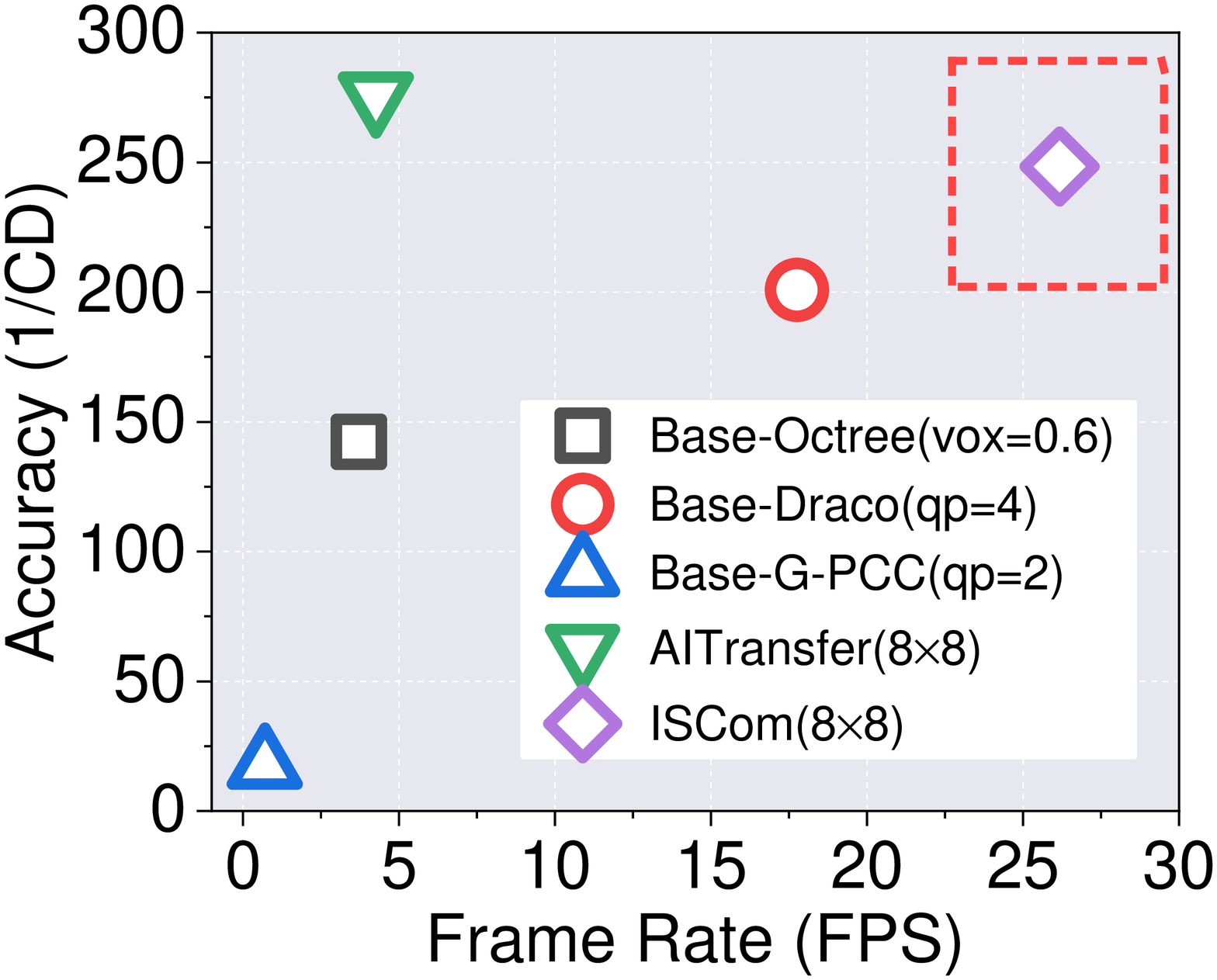}}
		\subfigure[Accuracy and FPS in WiFi]{
			\includegraphics[width=4.3cm]{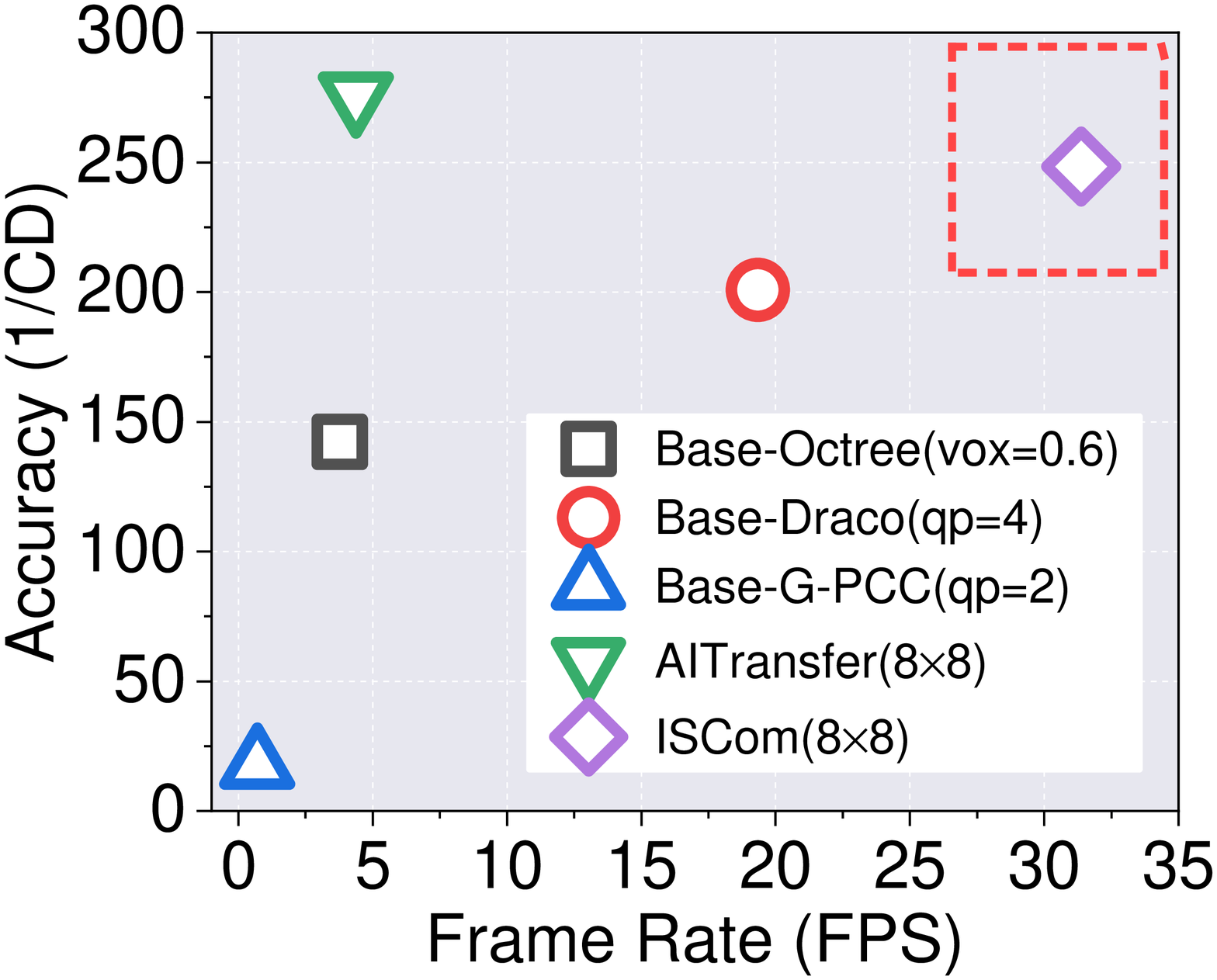}}
		\subfigure[Accuracy and FPS in 5G]{
			\includegraphics[width=4.3cm]{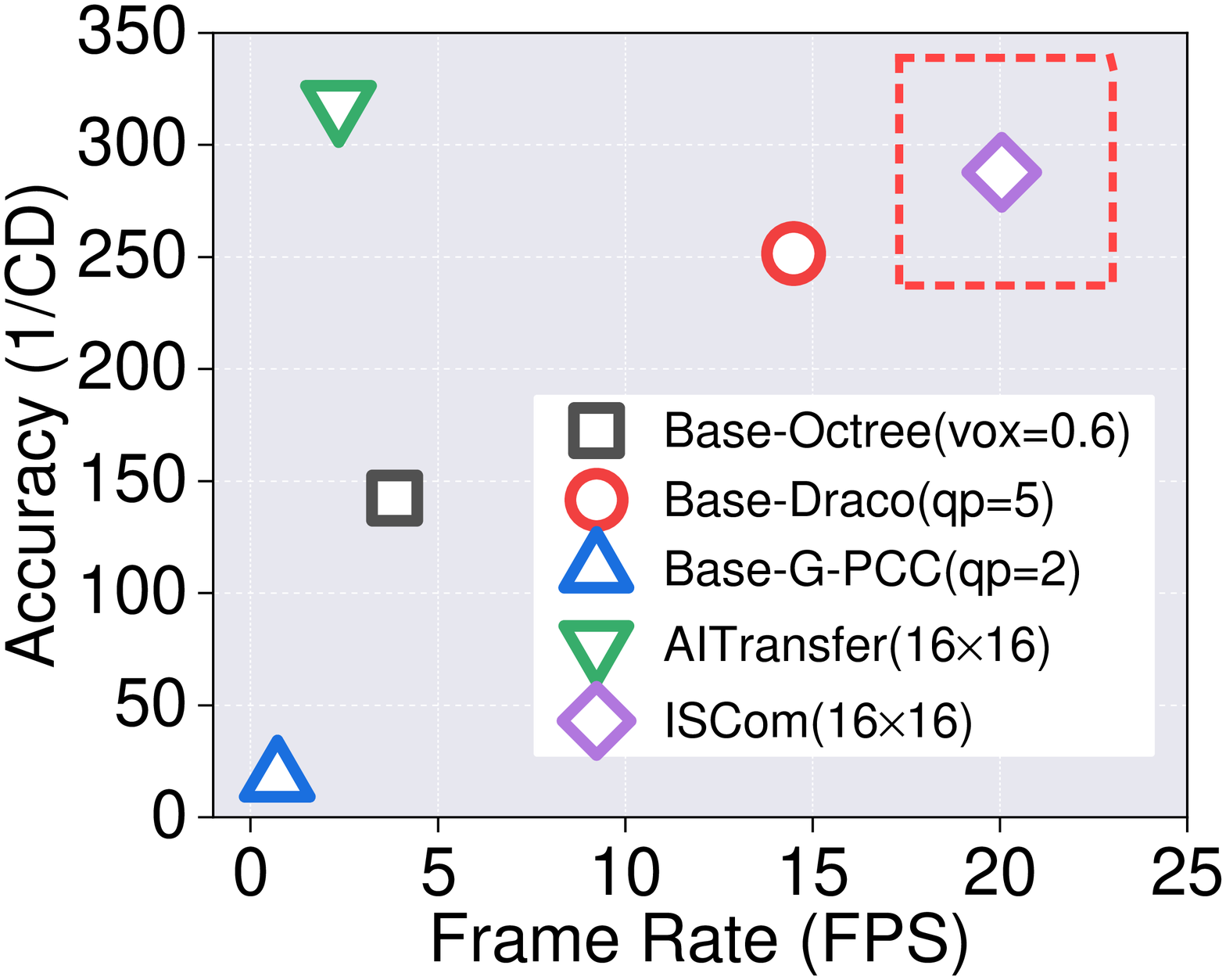}}
		\caption{Results of accuracy and frame rate on \textit{device-1}}
		\label{Fig5211}
	\end{minipage}%
	\vspace{-0.2cm}
\end{figure*}

\begin{figure*}[!htb]
	\begin{minipage}[t]{\linewidth}
		\centering
		\subfigure[Accuracy and FPS in 3G]{
			\includegraphics[width=4.3cm]{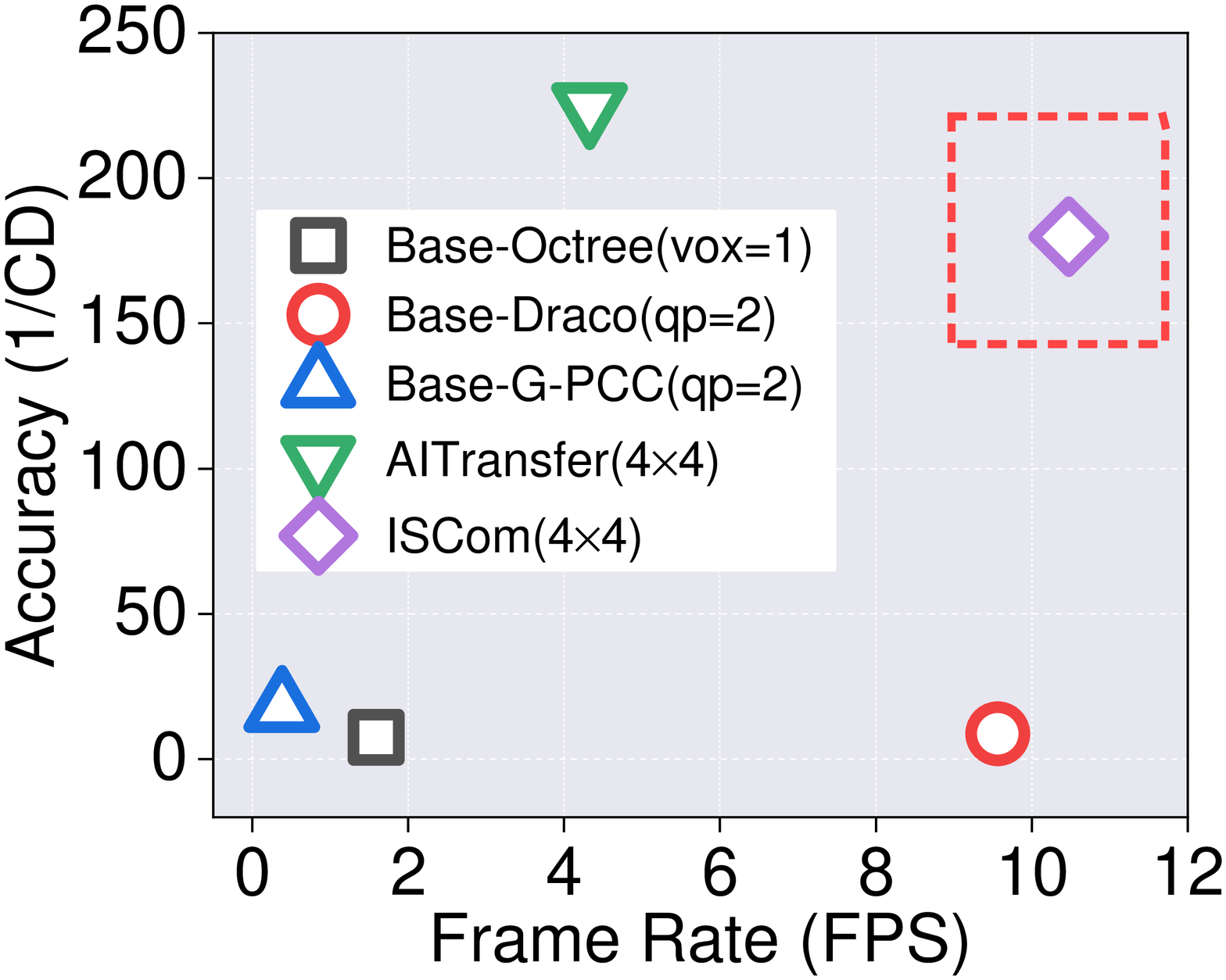}}
		\subfigure[Accuracy and FPS in 4G]{
			\includegraphics[width=4.3cm]{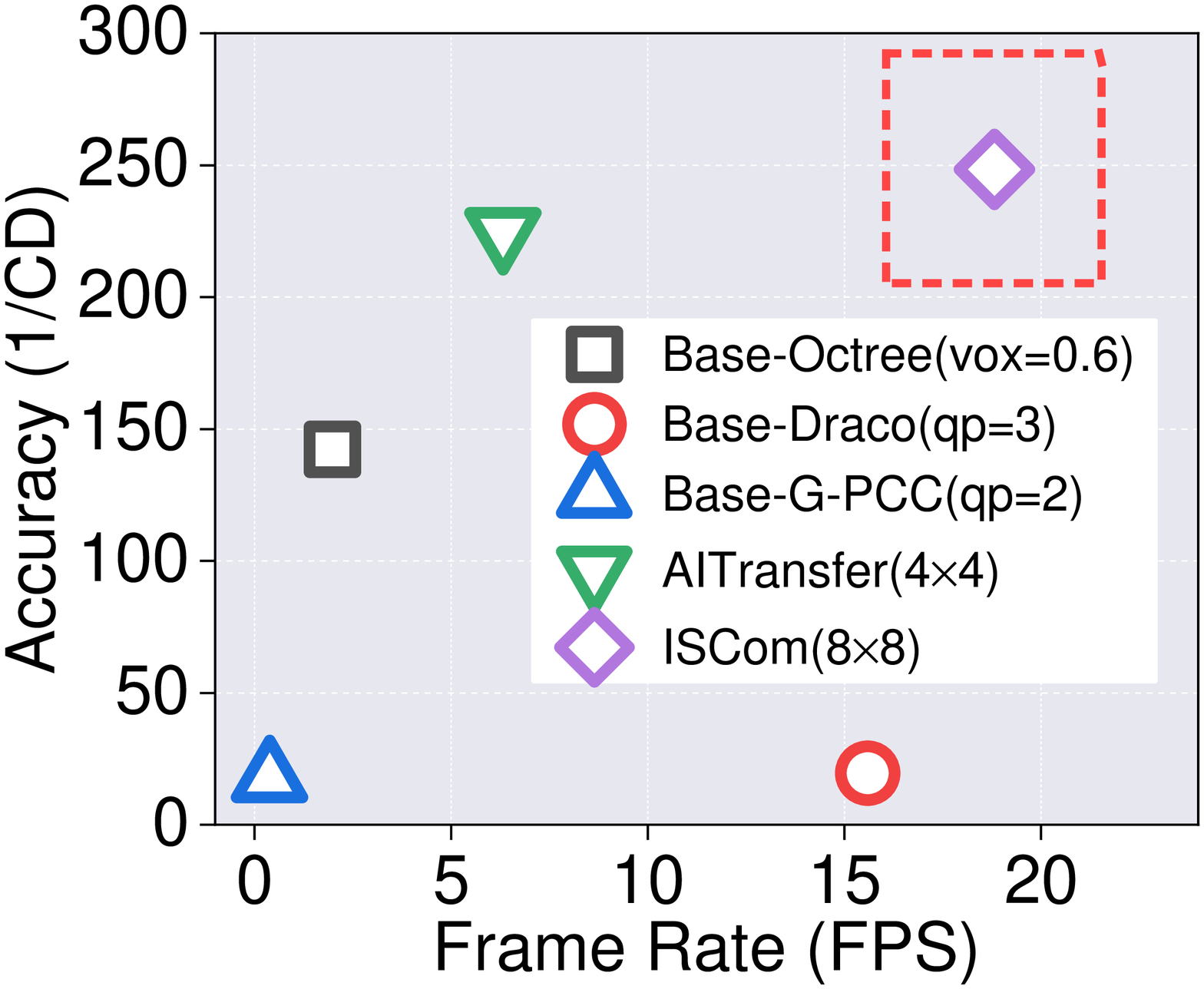}}
		\subfigure[Accuracy and FPS in WiFi]{
			\includegraphics[width=4.3cm]{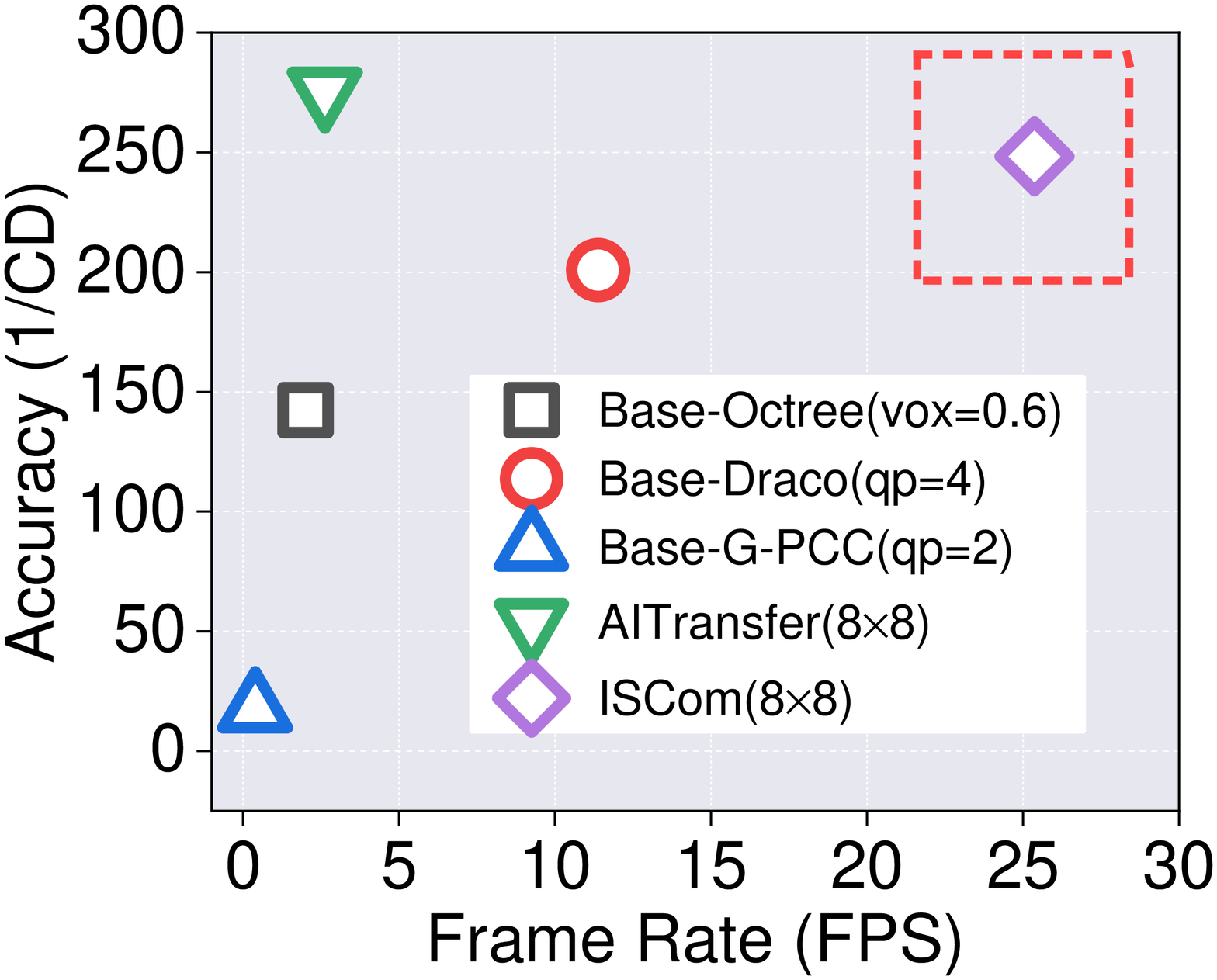}}
		\subfigure[Accuracy and FPS in 5G]{
			\includegraphics[width=4.3cm]{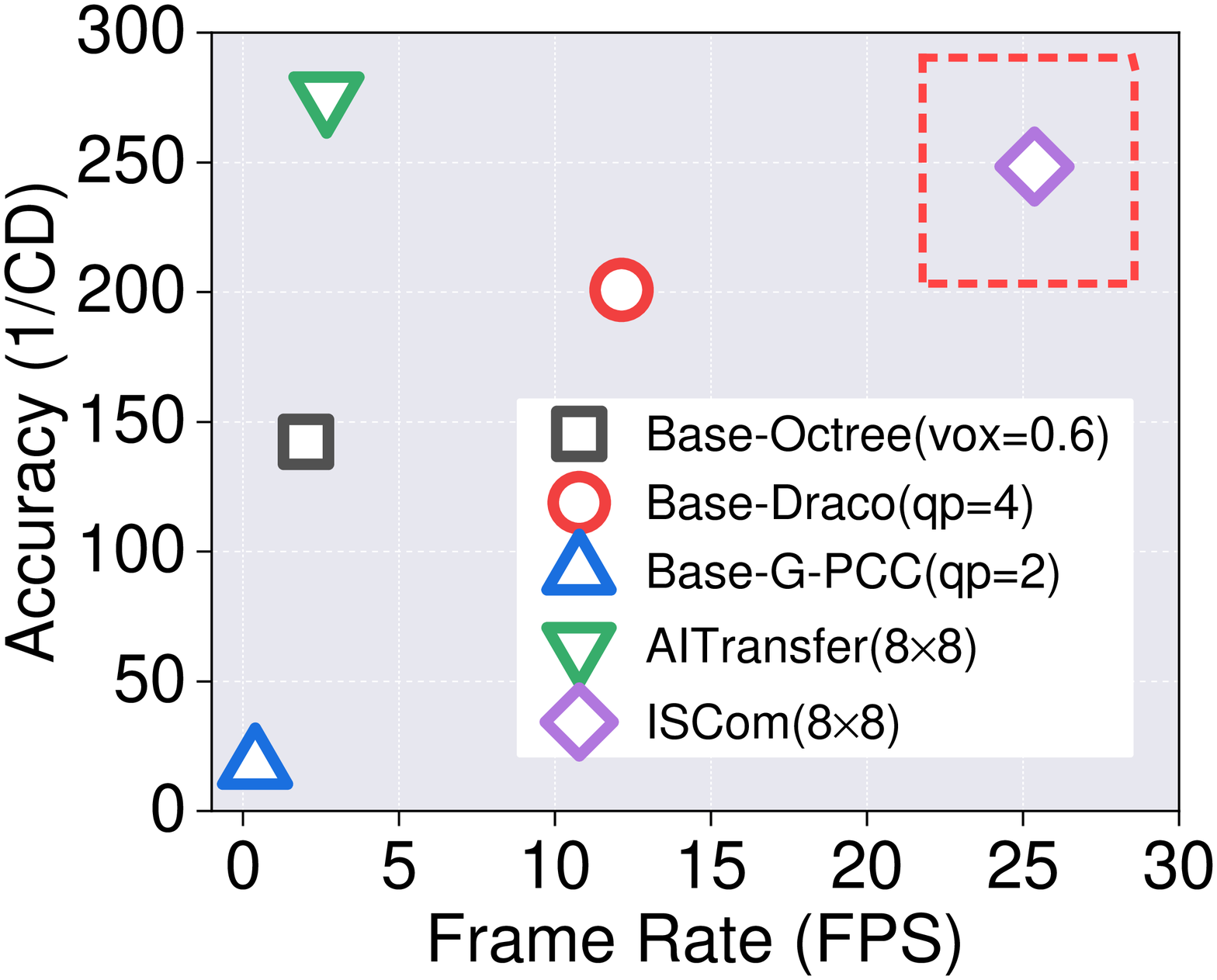}}
		\caption{Results of accuracy and frame rate on \textit{device-2}}
		\label{Fig5212}
	\end{minipage}%
	\vspace{-0.2cm}
\end{figure*}

\begin{figure*}[!htb]
	\begin{minipage}[t]{\linewidth}
		\centering
		\subfigure[Accuracy and FPS in 3G]{
			\includegraphics[width=4.3cm]{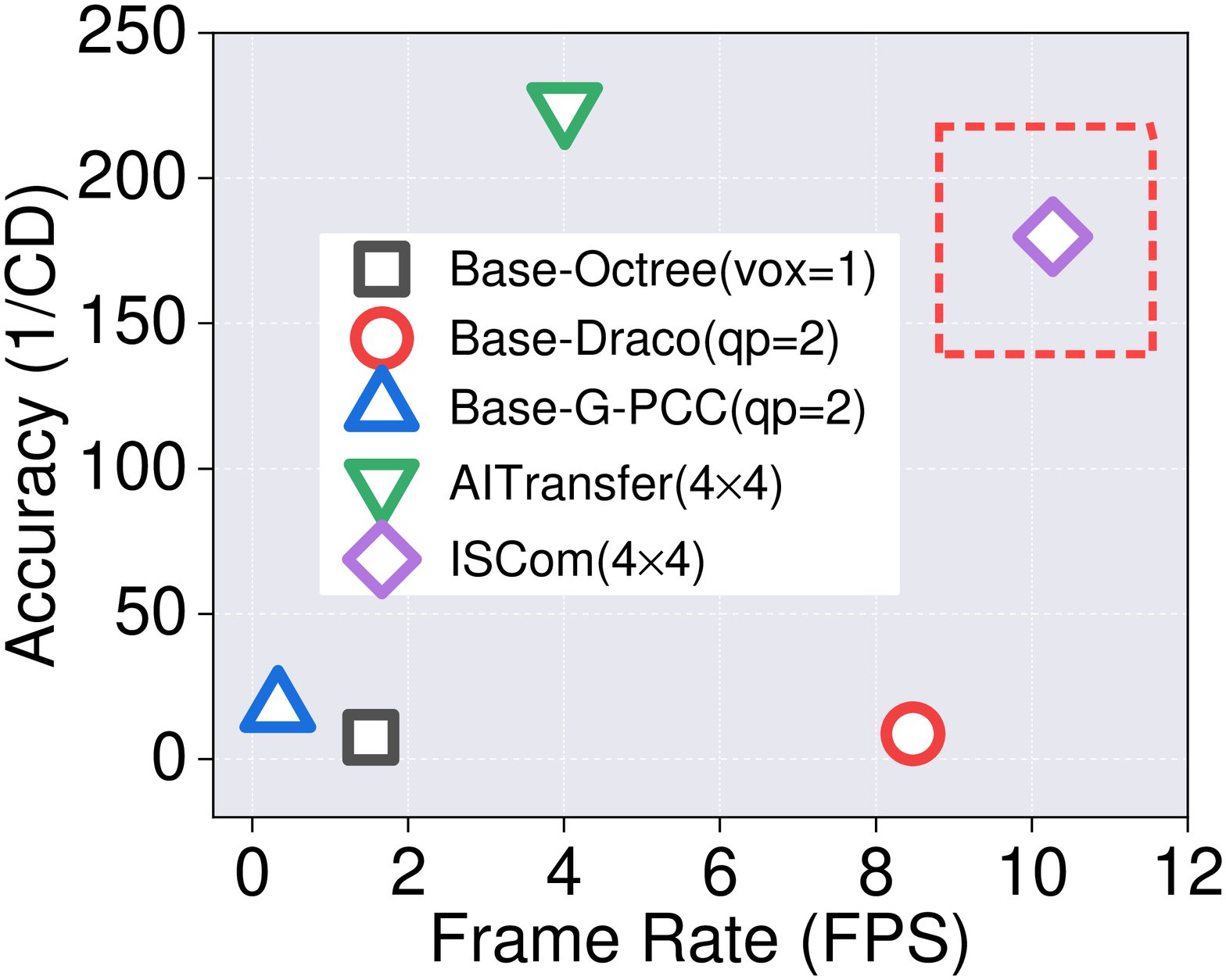}}
		\subfigure[Accuracy and FPS in 4G]{
			\includegraphics[width=4.3cm]{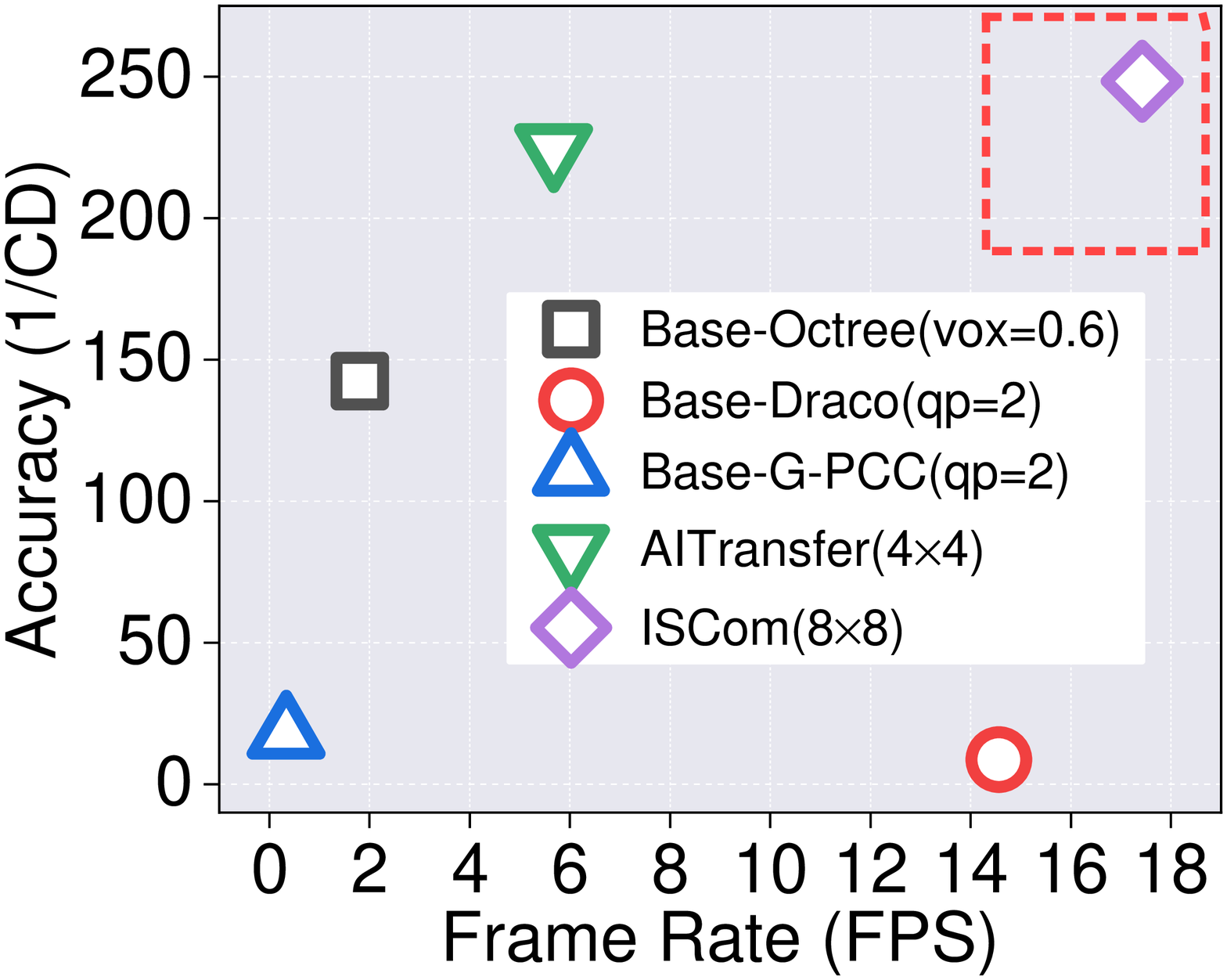}}
		\subfigure[Accuracy and FPS in WiFi]{
			\includegraphics[width=4.3cm]{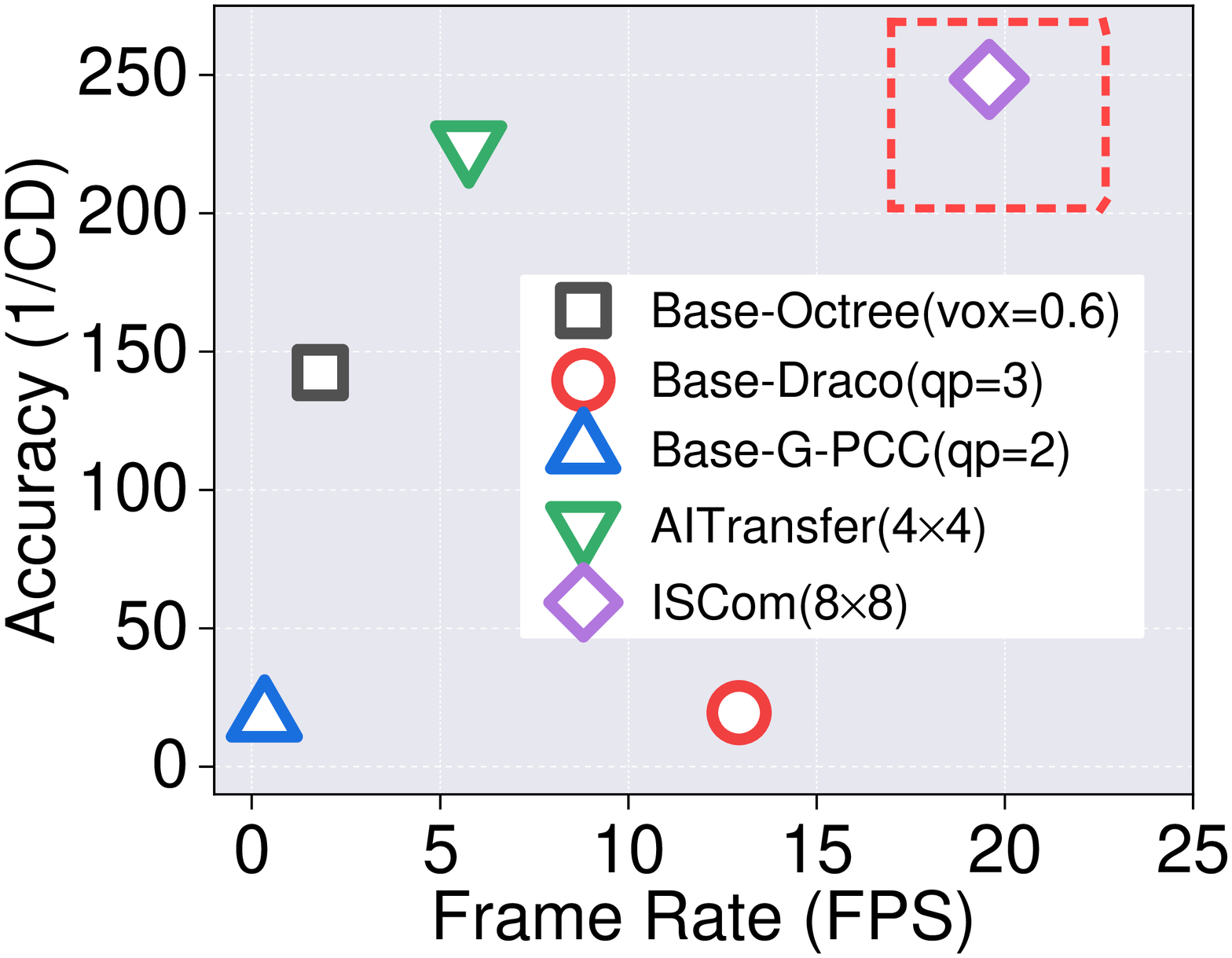}}
		\subfigure[Accuracy and FPS in 5G]{
			\includegraphics[width=4.3cm]{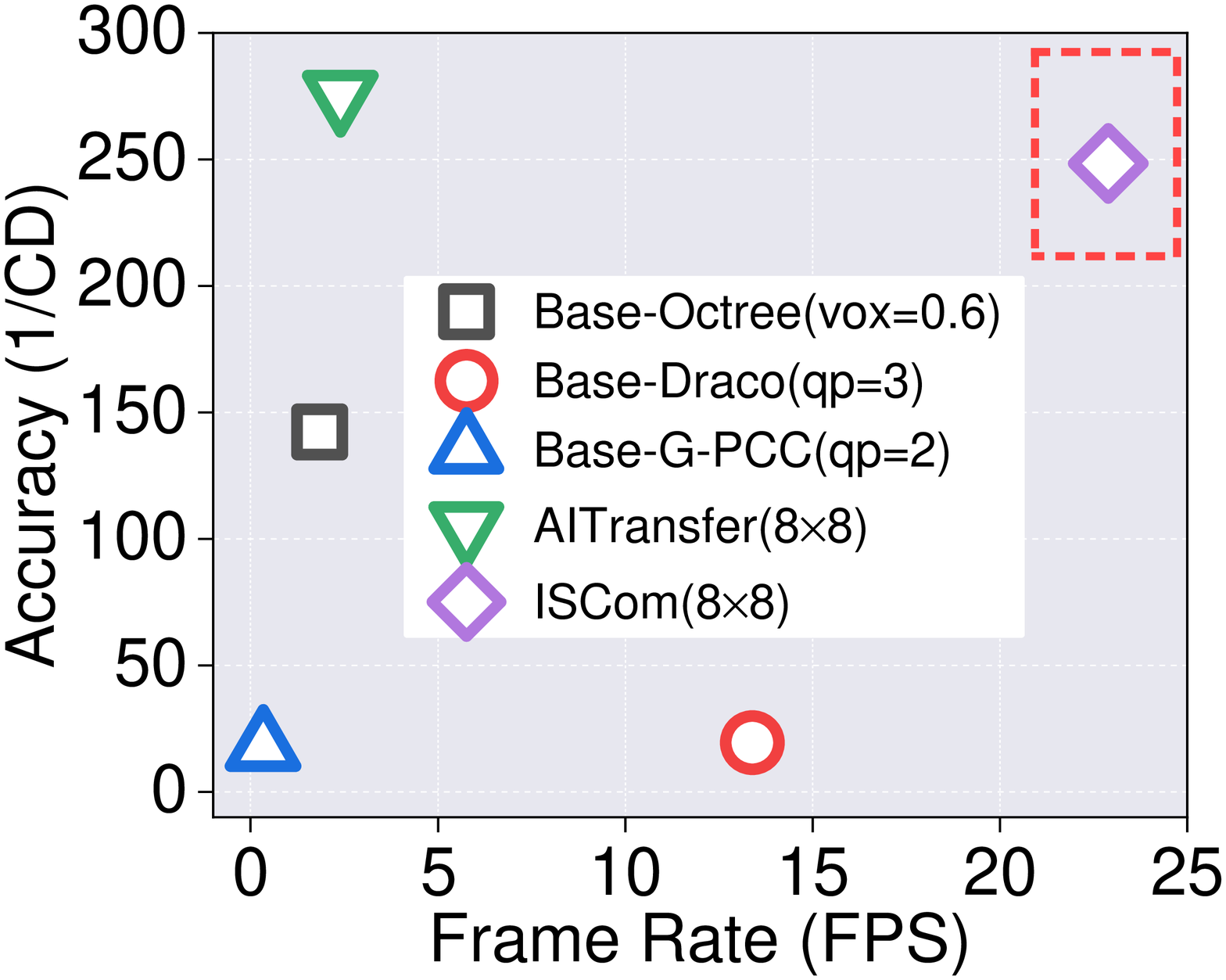}}
		\caption{Results of accuracy and frame rate on \textit{device-3}}
		\label{Fig5213}
	\end{minipage}%
	\vspace{-0.2cm}
\end{figure*}

We evaluate the received video quality in ISCom and alternative systems.
Specifically, we transmit the testing video without colour information under three different terminals and four network conditions (3G, 4G, WiFi, and 5G), respectively. 
For simplicity, we name the three different terminals as: \textit{device-1} (equipped with a 2.92~GHz CPU), \textit{device-2} (equipped with a 2.3~GHz CPU), and \textit{device-3} (equipped with a 2.2~GHz CPU).
The testing video contains 200 point cloud frames, where each frame is about 3.8 MB.
We show the results of (1/CD) and frame rate in Figure~\ref{Fig521HD}-\ref{Fig5213} performance, where the frame rate is jointly determined by the transmission and decoding phases.

We conclude that: 
(1)~With better network and device conditions, the constructed video accuracy and frame rate are becoming higher.
Moreover, to make the frame rate of all methods comparable in an order of magnitude, we have dynamically adjusted the parameters of each comparison system.
(2)~Base-G-PCC achieves the worst frame rate and quality. 
The reason is that G-PCC uses time-consuming compression algorithms in exchange for a larger compression ratio. 
To keep the frame rate at the same order of magnitude, we select a smaller depth parameter, which leads to lower video quality. 
(3)~The reconstructed accuracy in Base-Octree is related to the voxel size setting.
Although selecting a smaller voxel size such as 0.6 improves accuracy when compared with selecting a voxel size of 1, 
Base-Octree still fails to achieve high frame rates due to time-consuming encoding and decoding.
(4)~Draco is an efficient compression method, so Base-Draco achieves high frame rates. However, Draco ignores the accuracy, especially under a lower depth setting.
(5)~AITransfer almost achieves the best video quality, 
it can also obtain a higher compression ratio~\cite{huang2021aitransfer} benefiting from an AI-powered mechanism.
However, the mechanism based on deep learning is resource intensive for mobile devices.
Thus, AITransfer results in extremely low frame rates.
(6) Although in terms of accuracy, ISCom is inferior to AITransfer, especially on device 1, ISCom achieves overall better accuracy and frame rates.
To sum up, ISCom can achieve a balanced trade-off and obtain the most stable performance in all possible environments.
Moreover, we show two other representative results of (1/HD) and frame rate in Figure~\ref{Fig521HD}, in which the results are similar to that of CD.


\subsubsection{\textbf{Evaluation of streaming efficiency}}
\begin{figure}[htbp]
	\begin{minipage}[t]{\linewidth}
		\centering
		\subfigure[Streaming performance in 3G]{
			\includegraphics[width=8.6cm]{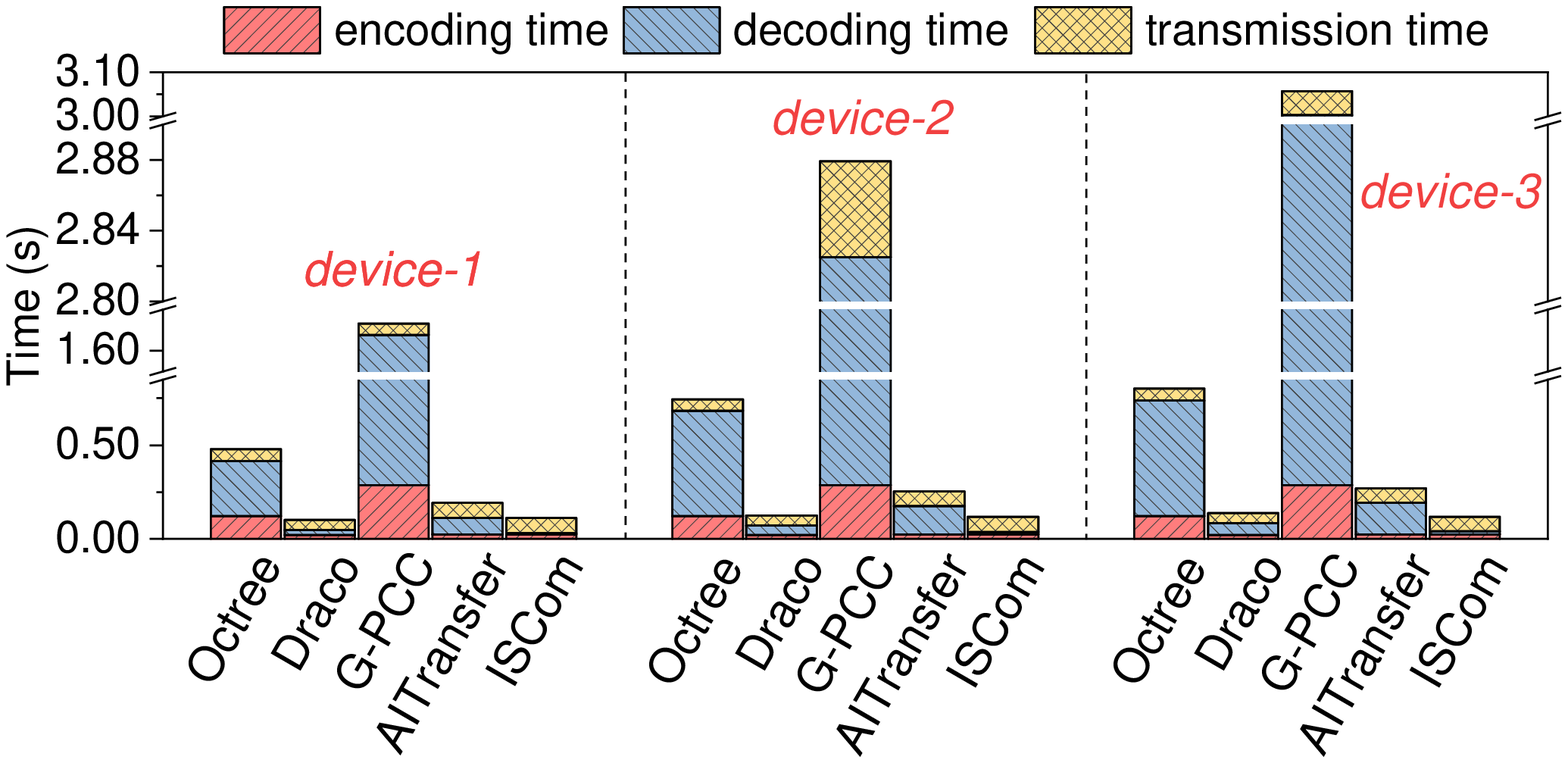}}
		\subfigure[Streaming performance in 4G]{
			\includegraphics[width=8.6cm]{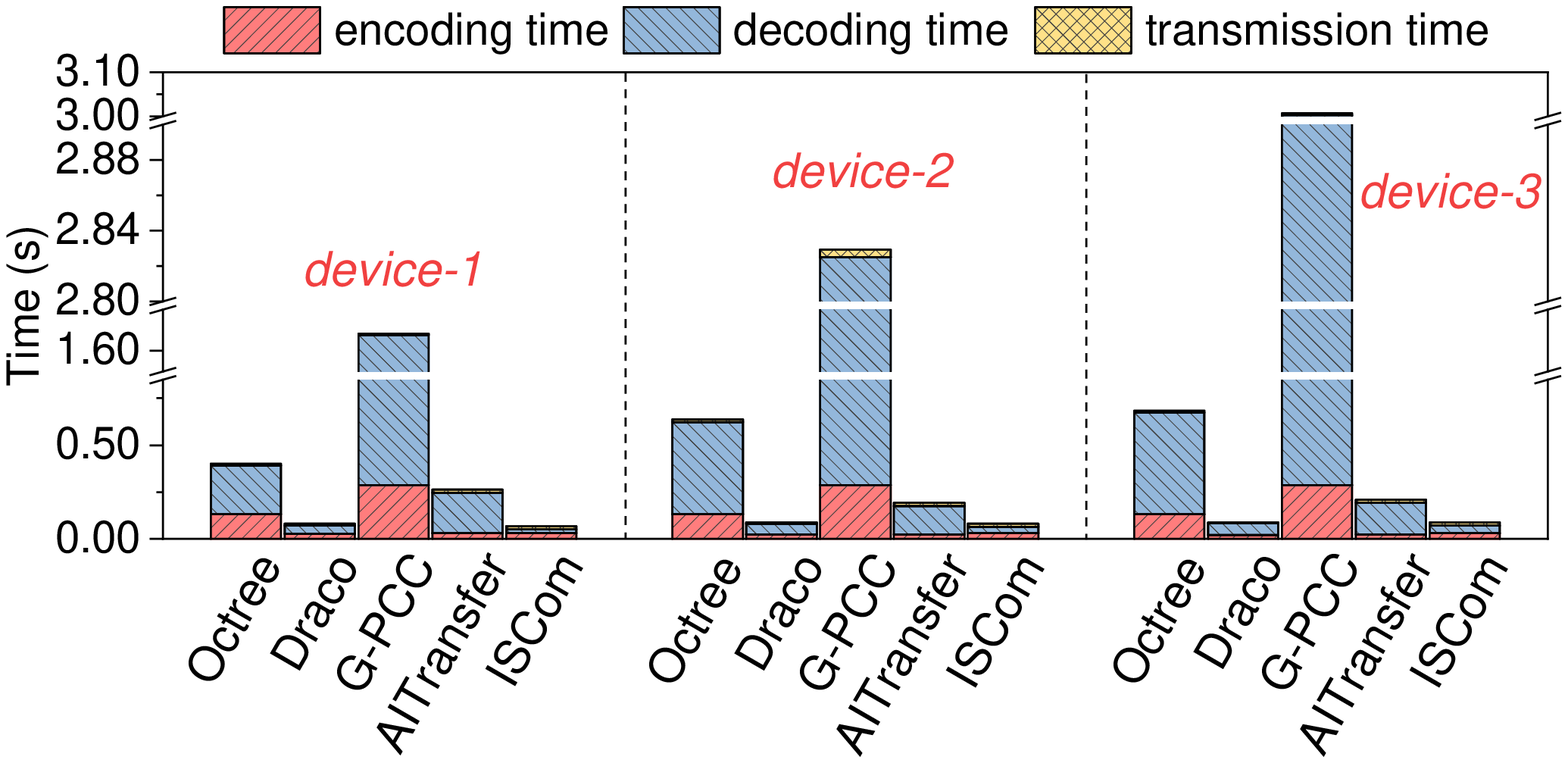}}
		\subfigure[Streaming performance in WiFi]{
			\includegraphics[width=8.6cm]{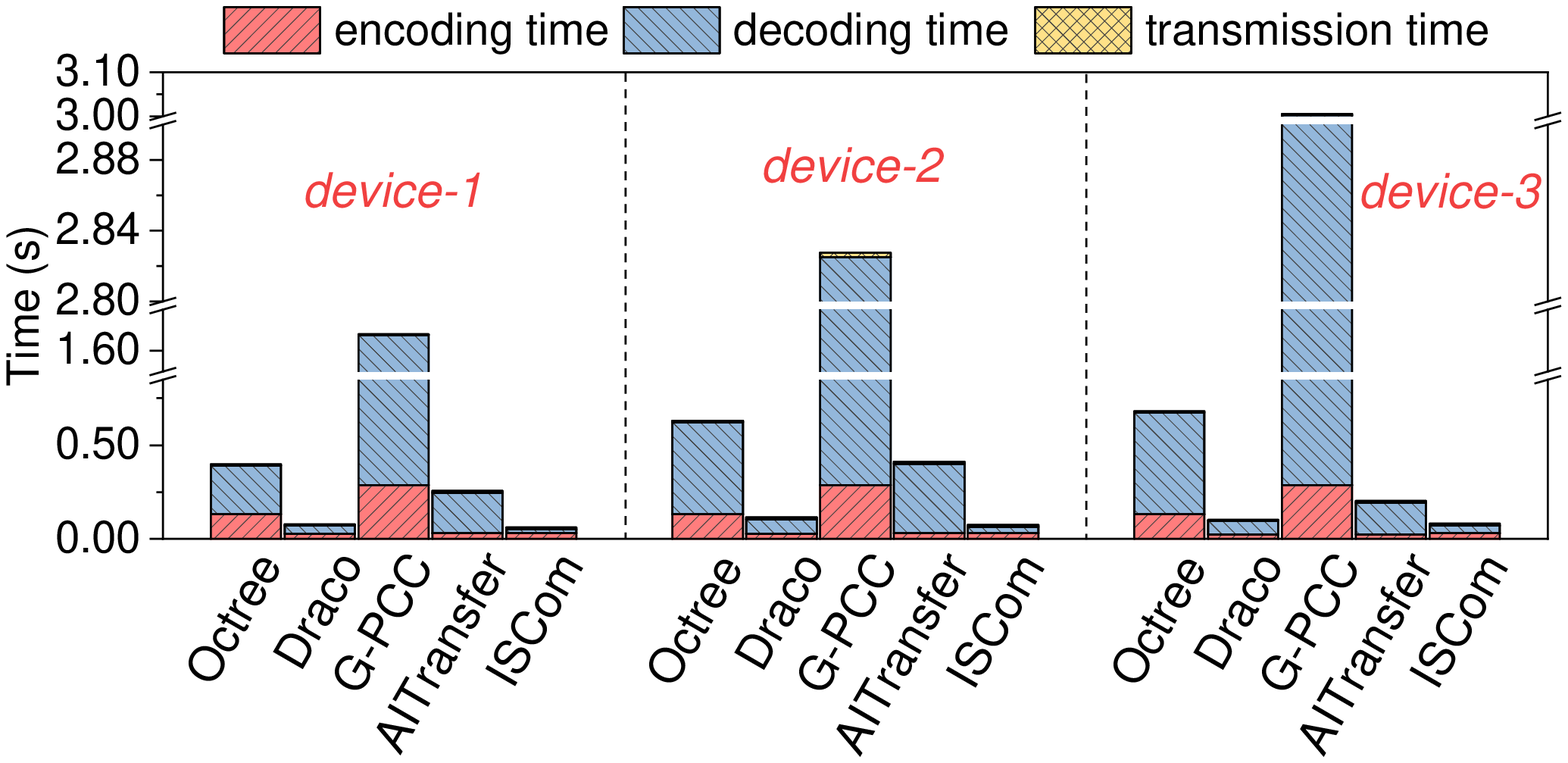}}
		\subfigure[Streaming performance in 5G]{
			\includegraphics[width=8.6cm]{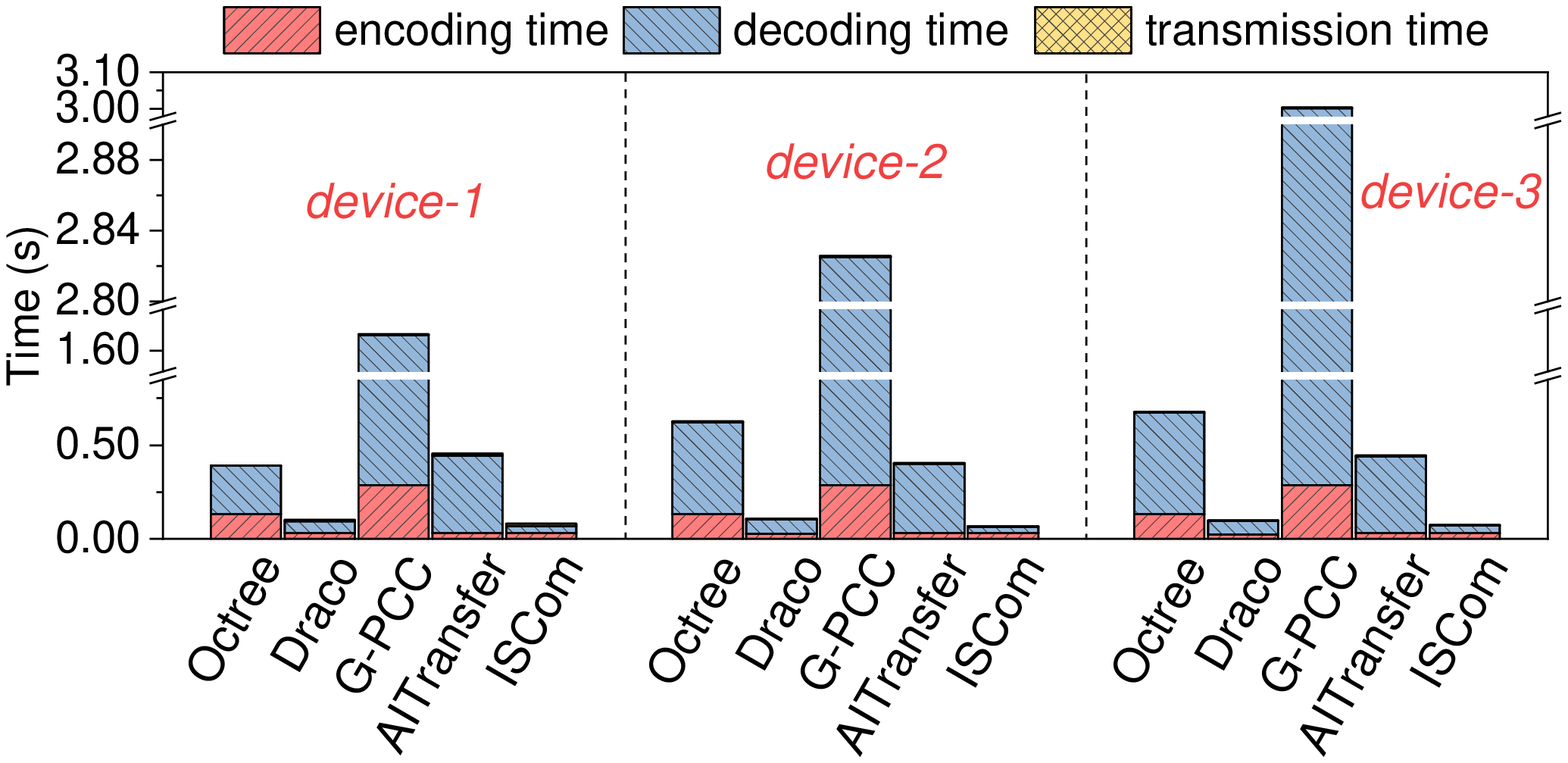}}
		\caption{The results of streaming efficiency}
		\label{Fig522}
	\end{minipage}%
	\vspace{-0.2cm}
\end{figure}

We evaluate the streaming efficiency of ISCom and alternative systems from three phases, i.e., encoding, transmission, and decoding.
We show the average times of each phase under different network conditions and devices in Figure~\ref{Fig522}.
The results illustrate that:
(1)~With the network conditions ranging from 3G to 5G, the transmission time in all systems becomes shorter.
Also, as the computing power of the device becomes stronger, the decoding time in all systems becomes shorter.
(2)~For all the systems, the decoding time takes the longest compared to encoding and transmission, which illustrates that decoding point cloud videos challenges mobile devices and becomes the bottleneck when improving efficiency. 
Moreover, G-PCC has the longest decoding time, and this phenomenon is consistent with the last experiment. 
(3)~ISCom achieves the shortest time in terms of encoding, decoding, and transmission in all kinds of environments, which demonstrates its streaming efficiency. 

\subsubsection{\textbf{Evaluation of runtime overhead}}
We evaluate the runtime overhead of five systems.
Specifically, we monitor the average CPU usage, memory usage (Mem), temperature (Temp), and frame rate (FPS), respectively, when transmitting a test video that has the same quality.
Since the results in different environments are comparable, we just present the results of the 5G network and \textit{device-1} in Table~\ref{tab1}.

\begin{table}[htbp]
\caption{The results of runtime overhead}
\centering
\renewcommand\arraystretch{1.4}
\resizebox{0.5\textwidth}{!}{%
\begin{tabular}{|l|c|c|c|c|}
\hline
Methods               & CPU (\%) & Mem (MB) & Temp ($^{\circ}$C) & FPS \\ \hline
Base-Octree(vox=0.6) & 27.41    & 97.57       & 44.56              & 3.85             \\
Base-Draco(qp=5)     & 20.80    & 68.73       & 43.96              & 14.49            \\
Base-G-PCC(qp=2)     & 16.64    & 76.74       & 43.31              & 0.71             \\
AITransfer(16x16)     & 14.51    & 291.85      & 47.21              & 2.35             \\
ISCom(16x16)          & 7.62     & 170.14      & 42.14              & 20.05            \\ \hline
\end{tabular}
}
\label{tab1}
\end{table}

Table~\ref{tab1} illustrates that:
(1) ISCom achieves the lowest average CPU usage and temperature while keeping the highest frame rate, which is 47.5\% and 2.7\% better than the next best system, respectively.
(2) AI-driven systems requires more memory consumption than traditional compression-based systems, this is due to the fact that the memory needs to store the input data, weight parameters, and activation function of the deep neural network to calculate the error gradient. 
However, ISCom reduces memory usage by 41.7\% compared to AITransfer, this is because we introduce the pruning and quantitative methods in ISCom's end-to-end model,
which demonstrates the effectiveness of our designed lightweight network.

\subsection{In-depth Analysis}
\subsubsection{\textbf{Improvements of Interest-aware module}}

To verify the effectiveness of the designed interest-aware module, we conduct an ablation study to compare ISCom with the ISCom without the interest-aware module, named ISCom-NoROI.
We conduct this experiment using ISCom ($16\times16$) model when using a 5G network on \textit{device-1}.
Table~\ref{tab2} presents the time of each phase.
The results show that the input video size of ISCom-NoROI is about 30 times larger than ISCom, and thus the time consumption of encoding, transmission, and decoding is also 30 times longer than that of ISCom. This is because ISCom can adaptively detect the user's ROI, and then cull the objects and backgrounds that the user may not care about.
This interest-aware transmission mechanism greatly reduces the transmitted data volume.
On the other hand, employing such a mechanism inevitably introduces additional computational latency, including coarse-grained and fine-grained ROI selection.
However, the operation of ROI selection is usually conducted on high-performance multimedia servers, it is feasible to increase some burden on the server side to bring significant benefits to the resource-constrained terminals.

\begin{table}[htbp]
\centering
\caption{The improvements of interest-aware module}
\label{tab2}
\renewcommand\arraystretch{1.3}
\resizebox{0.4\textwidth}{!}{%
\begin{tabular}{|l|c|c|}
\hline
Metrics               & ISCom & ISCom-NoROI \\ \hline
Input size (MB)       & 1.100 & 34.400      \\ \hline
Encoding time (s)     & 0.018 & 0.556       \\ \hline
Transmission time (s) & 0.029 & 2.752       \\ \hline
Decoding time (s)     & 0.011 & 0.335       \\ \hline
\end{tabular}
}
\end{table}

\subsubsection{\textbf{Improvements of lightweight decoder}}
To verify the effectiveness of the lightweight decoder, we compare ISCom with ISCom which only conducts pruning operations, named ISCom-P.
We also compare them with AITransfer that do not conduct any lightweight operation. From the results in Table~\ref{tab3}, we can observe that: 
(1) For each method, the smaller the compression ratio, the smaller the value of CD and HD, which illustrates that the reconstructed accuracy is negatively correlated with the compression ratio. The reason is that more semantic information is used during the feature extraction of ISCom.
(2) After the pruning operation, ISCom-P reduces some accuracy compared with AITransfer, and when the compression ratio becomes large, this reduced accuracy is also larger.
Besides, it is not surprising that the accuracy of ISCom further reduces due to the quantization operation compared with ISCom-P.
(3) The pruning enables ISCom-P and ISCom to achieve lower model sizes and inference times.
Moreover, ISCom conducts quantization operations to make its network model more lightweight, which leads to ISCom achieving about 10 times shorter than AITransfer, and about 3 times shorter than ISCom-P.

\begin{table*}[htbp]
\scriptsize
\centering

\caption{Results of lightweight decoder}
\renewcommand\arraystretch{1.4}
\resizebox{0.98\textwidth}{!}{%
\begin{tabular}{|l|ccc|ccc|ccc|}
\hline
                    & \multicolumn{3}{c|}{AITransfer}                                          & \multicolumn{3}{c|}{ISCom-P}                                & \multicolumn{3}{c|}{ISCom}                                               \\ \hline
                    & \multicolumn{1}{c|}{4x4}      & \multicolumn{1}{c|}{8x8}      & 16x16    & \multicolumn{1}{c|}{4x4}      & \multicolumn{1}{c|}{8x8}      & 16x16    & \multicolumn{1}{c|}{4x4}      & \multicolumn{1}{c|}{8x8}      & 16x16    \\ \hline
CD                  & \multicolumn{1}{c|}{0.004450} & \multicolumn{1}{c|}{0.003626} & 0.003144 & \multicolumn{1}{c|}{0.005060} & \multicolumn{1}{c|}{0.003916} & 0.003337 & \multicolumn{1}{c|}{0.005562} & \multicolumn{1}{c|}{0.004026} & 0.003474 \\ \hline
HD                  & \multicolumn{1}{c|}{0.033654} & \multicolumn{1}{c|}{0.027324} & 0.025230 & \multicolumn{1}{c|}{0.042000} & \multicolumn{1}{c|}{0.030115} & 0.025475 & \multicolumn{1}{c|}{0.043795} & \multicolumn{1}{c|}{0.030480} & 0.028378 \\ \hline
Model Size (MB)     & \multicolumn{1}{c|}{3.10}     & \multicolumn{1}{c|}{5.20}     & 9.40     & \multicolumn{1}{c|}{0.52}     & \multicolumn{1}{c|}{1.80}     & 1.93     & \multicolumn{1}{c|}{0.37}     & \multicolumn{1}{c|}{0.65}     & 0.80     \\ \hline
Inference Time (ms) & \multicolumn{1}{c|}{87.50}    & \multicolumn{1}{c|}{217.00}   & 352.90   & \multicolumn{1}{c|}{29.15}    & \multicolumn{1}{c|}{65.40}    & 121.50   & \multicolumn{1}{c|}{9.60}     & \multicolumn{1}{c|}{21.32}    & 37.20    \\ \hline
\end{tabular}
}
\label{tab3}
\end{table*}

\subsubsection{\textbf{Analysis of DRL-based scheduler}}
We analyze the training convergence performance of the designed DRL scheduler and the analysis of the hidden layer of the policy network in Figure~12.
The experiment parameters are set as follows: training epoch is 500, the learning rate of policy and value networks is set to 0.005, penalty factor $\gamma$ is 0.88, and the number of hidden layers is 96.
Figure~12(a) presents the convergence performance of the reward design using different $\eta$.
The result shows that a large $\eta$ introduces a large convergence value, which means the scheduler selects the lightweight with a lower transmission cost.
A small $\eta$ presents a small reword convergence value, causing the scheduler selects the model with higher accuracy.
As $\eta$ increases, the convergence speed becomes slower due to the gradually increasing influence of transmission speed, the network bandwidth and the computing capability of the device. 
Thus, the DRL network requires training more epochs to achieve a better fitting.
Figure~12(b) evaluates the number of hidden layers in the policy network by setting $\eta$ to 0.5.
We observe that the reward has difficulty converging with fewer hidden layers, meaning the DRL is too simple to fit the complex scheduling.
This is illustrated by increasing hidden layers and presenting the reward convergence.
However, a large number of hidden layers causes an increase in the training difficulty and scheduling overhead.
 \begin{figure}[htbp]
	\begin{minipage}[t]{\linewidth}
		\centering
		\subfigure[Analysis of parameter $\eta$]{
			\includegraphics[width=0.48\textwidth]{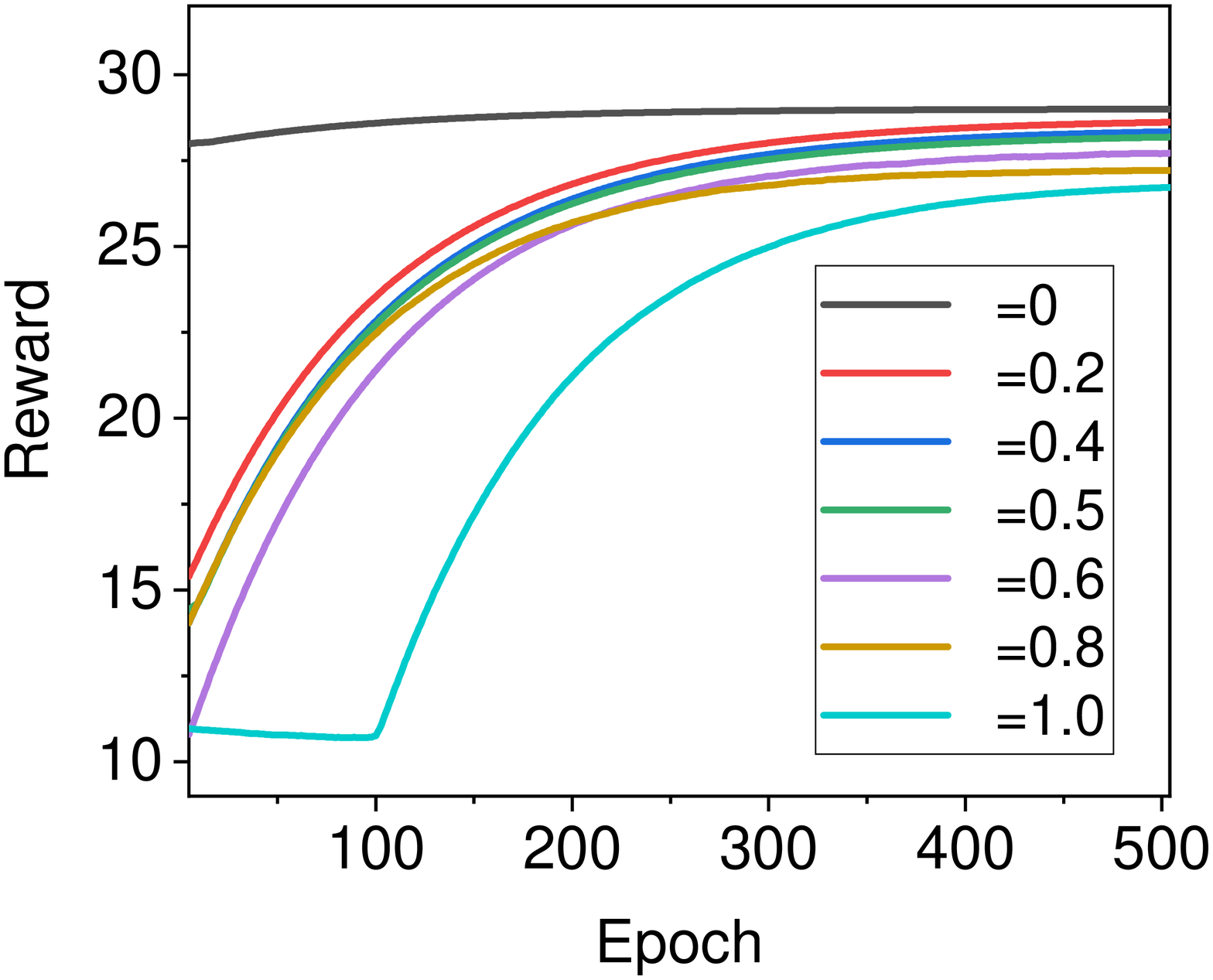}}
		\subfigure[Number of hidden layers]{
			\includegraphics[width=0.48\textwidth]{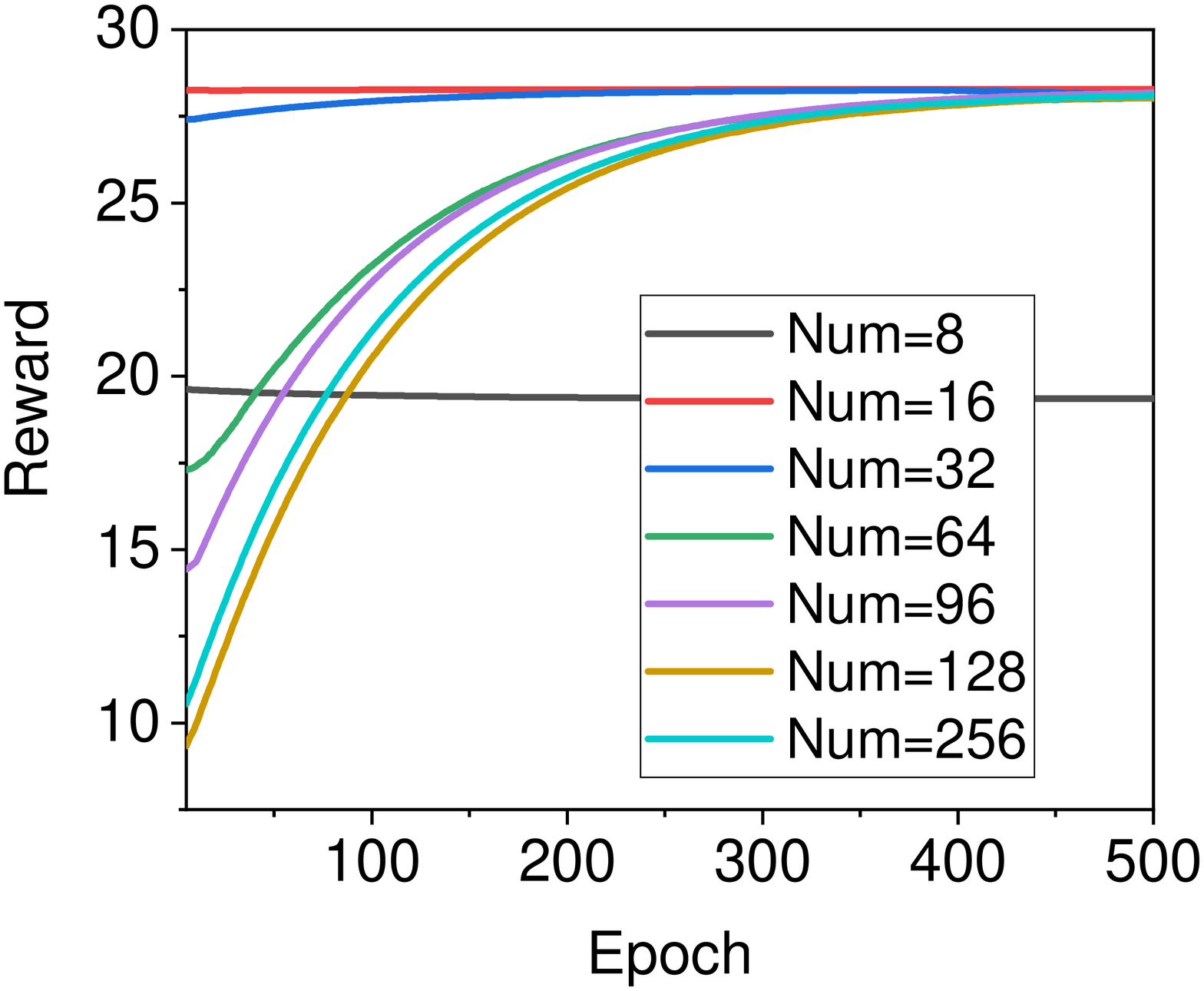}}
		\caption{Analysis of reward and policy network.}
		\label{Fig_12}
	\end{minipage}
	\vspace{-0.2cm}
\end{figure}

To verify the effectiveness of an intelligent DRL-based scheduler, we conduct an ablation study focusing on comparing two methods, i.e., ISCom with a DRL-based scheduler (ISCom (DRL)) and ISCom using a fixed encoder-decoder model (ISCom (16x16)).
We continuously transmit 300 frames of point cloud video with varying data volume under the environment of fluctuating network and computing capacity.
Table~\ref{tab4} present the results of these two methods. We can observe that ISCom (DRL) achieves a shorter average transmission time, maximum transmission time, average decoding time, and maximum decoding time than ISCom (16x16).
Therefore, ISCom (DRL) achieves a higher average frame rate and minimum frame rate than ISCom (16x16).
These results illustrate the necessity of adopting an intelligent model scheduler.
It is difficult to use a fixed encoder-decoder model in a dynamic environment, while the DRL-based scheduler makes scheduling according to the actual situation of the network and computing resources to improve streaming efficiency.
\begin{table}[!htbp]
\scriptsize
\centering
\caption{Results using DRL-based scheduler}
\label{tab4}
\renewcommand\arraystretch{1.3}
\resizebox{0.5\textwidth}{!}{%
\begin{tabular}{|l|c|c|}
\hline
Metrics                       & ISCom (DRL)     & ISCom (16x16) \\ \hline
Avg. transmission time (s)  & 0.0108  & 0.0130       \\
Max. transmission time (s) & 0.0131  & 0.0152       \\
Avg. decoding time (s)     & 0.0267  & 0.0301       \\
Max. decoding time (s)     & 0.0307  & 0.0493       \\
Avg. frame rate (fps)      & 26.5760 & 23.203       \\
Min. frame rate (fps)      & 22.8310 & 15.504       \\ \hline
\end{tabular}
}
\end{table}

\section{Related Work}
\subsection{Point Cloud Video Streaming}
Existing works stream point cloud video mainly using a combination of compression techniques~\cite{draco, graziosi2020overview,rusu20113d} and adaptive transmission mechanisms~\cite{hosseini2018dynamic,van2019towards,wang2021qoe,wang2022qoe}.
Compressing the point cloud video is the most intuitive way to reduce the transmitted data volume,
including geometry compression, attribute compression, and motion-compensated compression~\cite{krivokuca2019volumetric, chou2019volumetric}.
The geometry and attribute-based compression use octree or kd-tree structure to reduce spatial redundancy, such as PCL~\cite{rusu20113d}, Draco~\cite{draco}, G-PCC~\cite{graziosi2020overview}, GROOT~\cite{lee2020groot}, and deep learning-based codec~\cite{huang20193d,zhang2020mobile}.
The motion-compensated compression studies the relationship between adjacent frames to reduce time-domain redundancy, such as V-PCC~\cite{graziosi2020overview} and YuZu~\cite{zhang2022yuzu}.
On the other hand, 
the adaptive transmission mechanism optimises the resource utilisation by employing DASH standard~\cite{van2019towards}, tiling~\cite{li2020joint,liu2021point,subramanyam2020user,li2022optimal}, and viewport prediction~\cite{han2020vivo,lee2020groot}.
Our work is fundamentally different from these works, which use AI to extract the point cloud's semantic features for communications.

\subsection{Semantic Communications for Media Streaming}
Existing communication systems built on Shannon's theory ignore the semantic aspects of communication~\cite{shi2021semantic}.
To this end, Bao \textit{et al.}~\cite{bao2011towards} propose a model-theoretical approach for semantic data compression and communication.
Then, Shi \textit{et al.}~\cite{shi2021semantic} propose a federated edge intelligence architecture supporting resource-efficient semantic-aware networking. 
Recently, the paradigm of semantic communications has been applied to media streaming,  including text~\cite{xie2021deep}, audio~\cite{weng2021semantic}, image~\cite{zhang2022toward}, and 2D video~\cite{oquab2021low}.
For example, Oquab \textit{et al.}~\cite{oquab2021low} propose a video-chat compression which enables video calling only at a few kbits per second.
Thus, semantic communication is a promising paradigm for point cloud video streaming with extremely high bandwidth requirements. 
AITransfer~\cite{huang2021aitransfer} is the first system that streams the point cloud's semantic feature using an end-to-end neural network.
However,  when video streaming introduces the idea of semantic communications with AI inferring, it needs a higher computing requirement. 
This will change the trade-off from the communication-storage paradigm to the communication-storage-computation three-way joint optimization.

\section{Conclusion}
The paper presents ISCom, an interest-aware semantic communication scheme for immersive point cloud video service. ISCom addresses the critical challenges for real-time point cloud video streaming on resource-constrained devices by designing an ROI selection module, a lightweight decoder network, and an intelligent scheduler for online adaptive point cloud video service.
The two-stage efficient ROI selection method provides users' interest content and significantly reduces the data volume. The lightweight decoder is the key contribution to enabling the real-time reconstruction of resource-constrained terminals, and an intelligent scheduler can provide an adaptive service for various terminals and dynamic environments.
Our experimentation show that ISCom outperforms the advanced AI-driven method by at least 10 FPS and up to 22 FPS with considerable streaming quality.

ISCom has several significant advantages.
First, ISCom adopts the idea of semantic communication and AI-native technology to reduce the data volume through the interest-aware framework and achieves a high compression ratio by transmitting key semantic features.
Second, ISCom employs a lightweight neural network and intelligent scheduler to provide real-time point cloud video streaming even on resource-constrained devices. This feature allows ISCom to be generalised in many ubiquitous resource-constrained devices.

ISCom has two limitations. First, ISCom adopts the strategy of offline training and online flexible deployment, so it needs to cache a large number of neural network models. Although these models have been simplified by lightweight techniques, numerous models will consume considerable memory usage, which brings challenges to resource-constrained terminals. Second, ISCom uses AI-native technology to compress the spatial features of point clouds but does not study the temporal features of point cloud frames. It is a promising way to extend current AI-driven transmission capacity by combing some temporal domain optimization, such as MPEG standard.

In future research, we will 1) investigate methodologies of balancing communication, computation, and storage in AI-driven point cloud video transmission; and 2) apply ISCom to future metaverse use cases to support scalable and immersive user interactions.

\section*{Acknowledgment}
This research was supported in part by the National Key R\&D Program of China under Grant 2018YFE0205503, in part by the Funds for International Cooperation and Exchange of NSFC under Grant 61720106007, in part by the Project funded by China Postdoctoral Science Foundation 2022TQ0047 and 2022M710465, and in part by Academy of Finland under projects 319670 and 326305.
The authors would thank Jacky Cao for his diligent proofreading of this article.
\vspace{-0.2cm}
\ifCLASSOPTIONcaptionsoff
  \newpage
\fi

\bibliographystyle{IEEEtran}
\bibliography{IEEEabrv,IEEEexample}
\vspace{-1cm}
\begin{IEEEbiography}[{\includegraphics[width=1in,height=1.25in,clip,keepaspectratio]{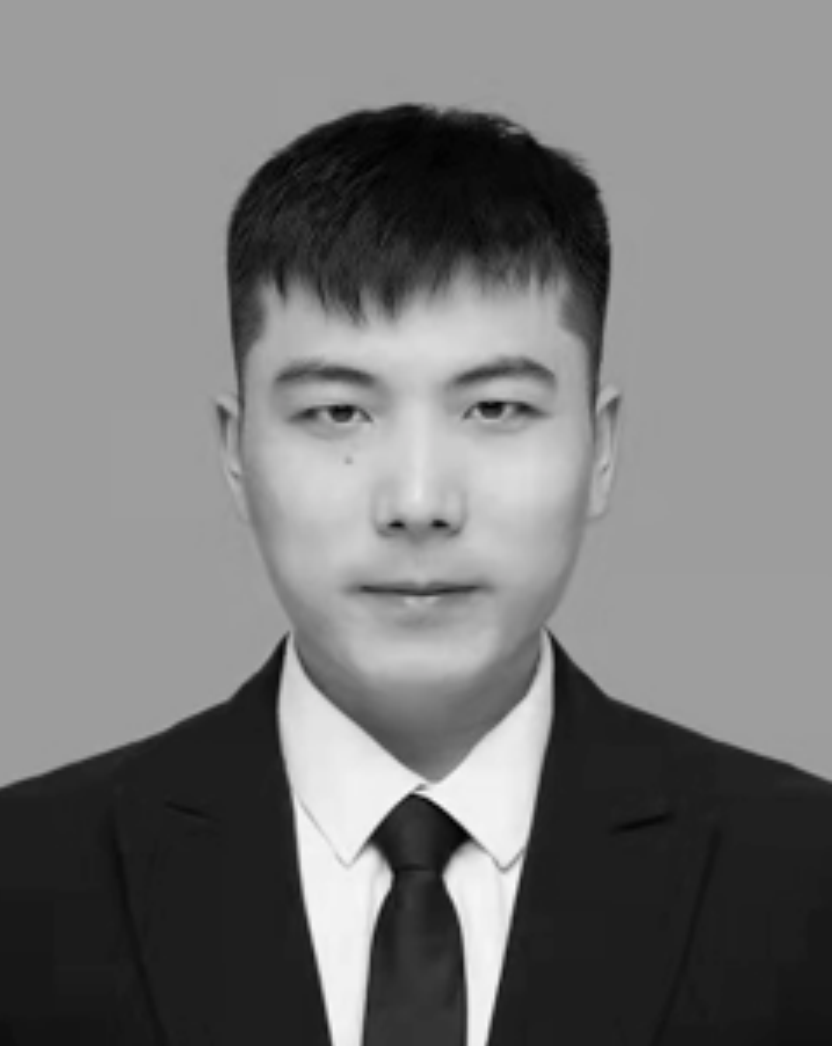}}]{Yakun Huang}
received the Ph.D degree in computer science from Beijing University of Posts and Telecommunications, in 2021. 
He has authored or co-authored over 10 technical papers in international journals and at conferences, including the IEEE Transactions on Mobile Computing, the IEEE Transactions on Service Computing, the IEEE Network, INFOCOM, ICDCS, MM. His current research interests include volumetric video streaming, mobile computing, edge computing, and distributed systems.	
\end{IEEEbiography}

\begin{IEEEbiography}[{\includegraphics[width=1in,height=1.25in,clip,keepaspectratio]{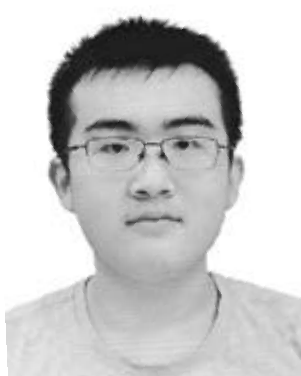}}]{Boyuan Bai}
is currently working toward a Ph.D. degree at the State Key Laboratory of Networking and Switching Technology, Beijing University of Posts and Telecommunications, Beijing, China. His current research interests include point clouds, video streaming transmission, and deep reinforcement learning.	
\end{IEEEbiography}

\begin{IEEEbiography}[{\includegraphics[width=1in,height=1.25in,clip]{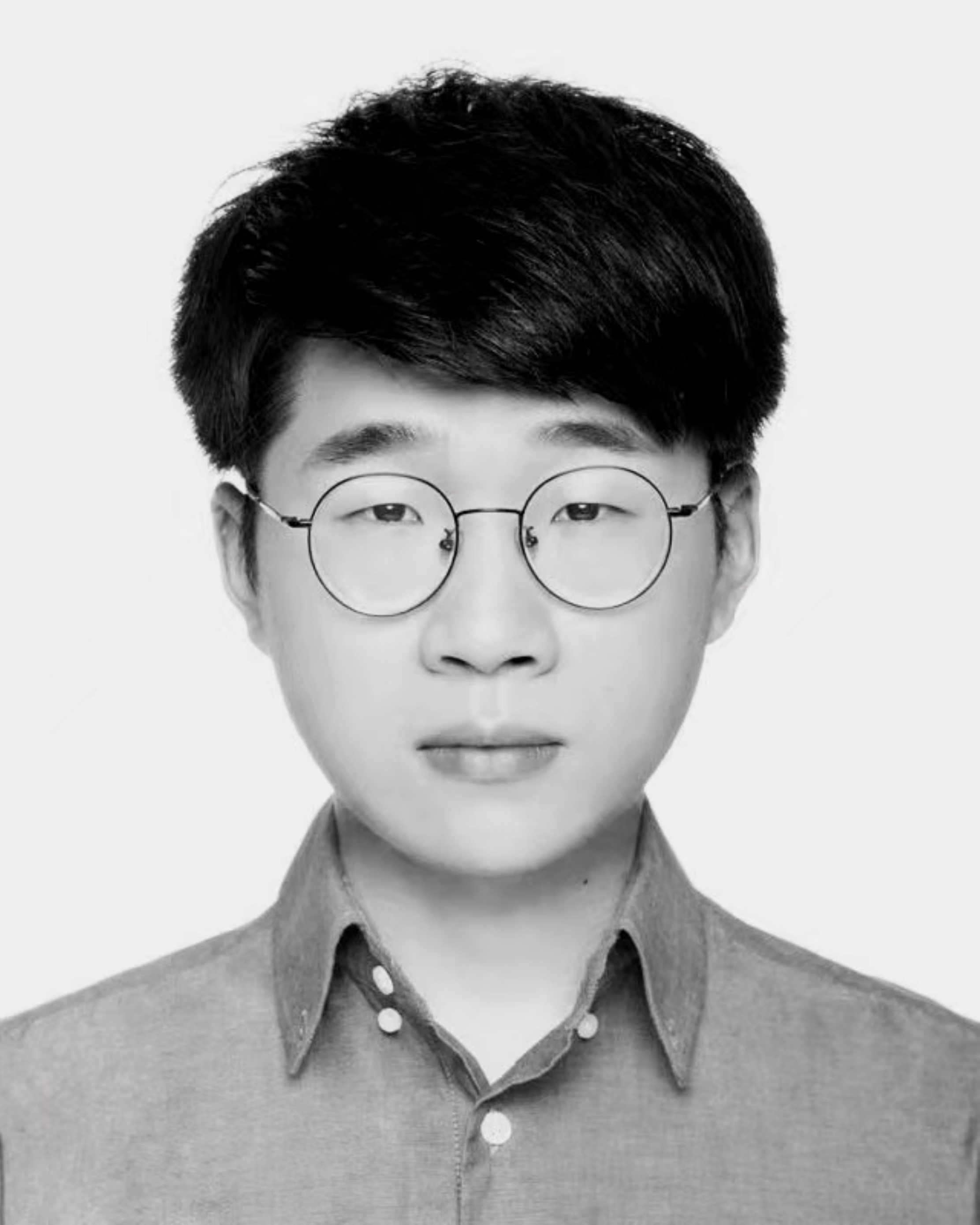}}]{Yuanwei Zhu}
is currently working toward a Ph.D. degree at the State Key Laboratory of Networking and Switching Technology, Beijing University of Posts and Telecommunications, Beijing, China. His current research interests include point clouds, video streaming transmission, and deep reinforcement learning.	
\end{IEEEbiography}

\begin{IEEEbiography}[{\includegraphics[width=1in,height=1.25in,clip,keepaspectratio]{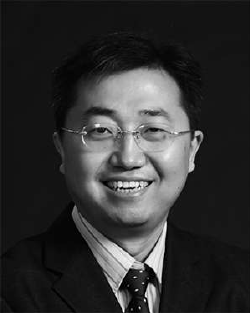}}]{Xiuquan Qiao}
is currently a Full Professor with the Beijing University of Posts and Telecommunications, Beijing, China, where he is also the Deputy Director of the Network Service Foundation Research Center, State Key Laboratory of Networking and Switching Technology. He has authored or co-authored over 60 technical papers in international journals and at conferences, including the IEEE Communications Magazine, Proceedings of IEEE, Computer Networks, IEEE Internet Computing, the IEEE TRANSACTIONS ON AUTOMATION SCIENCE AND ENGINEERING, and the ACM SIGCOMM Computer Communication Review. His current research interests include the future Internet, services computing, computer vision, distributed deep learning, augmented reality, virtual reality, and 5G networks. Dr. Qiao was a recipient of the Beijing Nova Program in 2008 and the First Prize of the 13th Beijing Youth Outstanding Science and Technology Paper Award in 2016. He served as the associate editor for Computing (Springer) and the editor board of China Communications Magazine.
\end{IEEEbiography}

\begin{IEEEbiography}[{\includegraphics[width=1in,height=1.25in,clip,keepaspectratio]{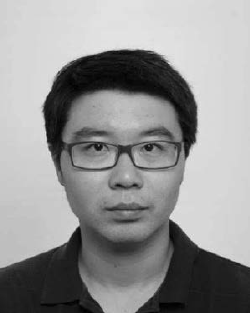}}]{Xiang Su}
is currently an Associate Professor with the Department of Computer Science, Norwegian University of Science and Technology, Norway, and the University of Oulu, Finland. He has extensive expertise in the Internet of Things, edge computing, mobile augmented reality, knowledge representations, and context modelling and reasoning. 
\end{IEEEbiography}

\begin{IEEEbiography}[{\includegraphics[width=1in,height=1.25in,clip,keepaspectratio]{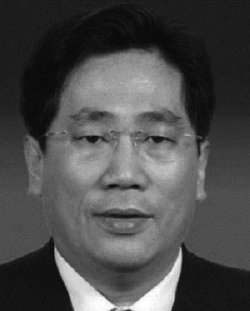}}]{Ping Zhang}
(Fellow, IEEE) received the M.S. degree in electrical engineering from Northwestern Polytechnical University, Xi’an, China, in 1986, and the Ph.D. degree in electric circuits and systems from the Beijing University of Posts and Telecommunications
(BUPT), Beijing, China, in 1990. He is currently a Professor with BUPT. His research interests include cognitive wireless networks, fourth-generation mobile communication, fifth-generation mobile networks, communications factory test instruments, universal wireless signal detection instruments, and mobile internet. He was the recipient of the First and Second Prizes from the National Technology Invention and Technological Progress Awards and First Prize Outstanding Achievement Award of Scientfic Research in College. He is the Executive Associate Editor-in-Chief of the Information Sciences of the Chinese Science Bulletin, the Guest Editor of IEEE Wireless Communications Magazine, and the Editor of China Communications.
\end{IEEEbiography}

\end{document}